\pdfoutput=1

\documentclass[twocolumn,superscriptaddress,aps,preprintnumbers,amsmath,amssymb,prd,nofootinbib]{revtex4}

\usepackage{epsfig}
\usepackage{amsmath}
\usepackage{amssymb}
\usepackage{subfigure}
\usepackage{cancel}
\usepackage{textcomp}
\usepackage{calc}
\usepackage{graphicx}
\usepackage{dcolumn}
\usepackage{color}
\usepackage{xcolor}
\usepackage{pifont}
\usepackage{slashed}
\usepackage{fdsymbol}
\usepackage{orcidlink}

\hypersetup{
colorlinks   = true, 
urlcolor     = red, 
linkcolor    = blue, 
citecolor   = blue 
}
\usepackage[capitalise]{cleveref}
\usepackage[normalem]{ulem} 
\definecolor{nicegreen}{rgb}{0.1,0.5,0.1}

\newcommand{\nn}{\nonumber}

\begin{document}


\preprint{CTPU-PTC-22-11}
\preprint{DESY-22-094}
\preprint{MS-TP-22-17}
\preprint{CERN-TH-2022-091}

\title{Axion dark matter from frictional misalignment}

\author{Alexandros Papageorgiou\orcidlink{0000-0002-2736-3026}} 
\email{papageo@ibs.re.kr}
\affiliation{Center for Theoretical Physics of the Universe, Institute for Basic Science\\
}
\author{Pablo Qu\'ilez\orcidlink{0000-0002-4327-2706}}
\email{pablo.quilez@desy.de}
\affiliation{Department of Physics, University of California, San Diego\\
}
\affiliation{Deutsches Elektronen-Synchrotron DESY\\
}

\author{Kai Schmitz\orcidlink{0000-0003-2807-6472}} 
\email{kai.schmitz@uni-muenster.de}
\affiliation{University of M\"unster, Institute for Theoretical Physics, 48149 M\"unster, Germany}
\affiliation{Theoretical Physics Department, CERN, 1211 Geneva 23, Switzerland}


\begin{abstract}
We study the impact of sphaleron-induced thermal friction on the axion dark-matter abundance due to the interaction of an axion-like particle (ALP) with a dark non-abelian gauge sector in a secluded thermal bath. Thermal friction can either enhance the axion relic density by delaying the onset of oscillations or suppress it by damping them. We derive an analytical formula for the \emph{frictional adiabatic invariant}, which remains constant along the axion evolution and which allows us to compute the axion relic density in a general set-up. Even in the most minimal scenario, in which a single gauge group is responsible for both the generation of the ALP mass and the friction force, we find that the resulting dark-matter abundance from the misalignment mechanism deviates from the standard scenario for axion masses  $m_a\gtrsim 100 \; {\rm eV}$. We also generalize our analysis to the case where the gauge field that induces friction and the gauge sector responsible for the ALP mass are distinct and their couplings to the axion have a large hierarchy as can be justified by means of alignment or clockwork scenarios. We find that it is easy to open up the ALP parameter space where the resulting axion abundance matches the observed dark-matter relic density both in the traditionally over- and underabundant regimes. This conclusion also holds for the QCD axion.
\end{abstract}


\date{\today}


\maketitle

\section{Introduction}

Axions are pseudo-Nambu-Goldstone bosons that arise in consequence of the spontaneous breaking of a chiral global $U(1)$ symmetry. Such a symmetry, if anomalous under $SU(3)_c$ in quantum chromodynamics (QCD),  may play an important role in resolving the strong CP problem \cite{Peccei:1977hh,Peccei:1977ur,Weinberg:1977ma,Wilczek:1977pj}. Nonetheless, the interest in axions extends way beyond the QCD axion; axions or ALPs arise in a variety of theories (e.g. ~\cite{Dienes:1999gw,Gelmini:1980re,Davidson:1981zd,Wilczek:1982rv,Cicoli:2013ana}). Furthermore, axions  are attractive dark-matter candidates, which can be nonthermally produced via the vacuum misalignment mechanism \cite{Preskill:1982cy,Abbott:1982af,Dine:1982ah,Higaki:2015jag,Higaki:2016yqk} or the decay of topological defects \cite{Kibble:1976sj,Kibble:1980mv,Davis:1985pt,Davis:1986xc,Harari:1987ht,Battye:1993jv,Higaki:2016jjh}. The former mechanism proceeds as follows. In the early Universe, the axion field is frozen at an initial field value due to Hubble friction. At later times, when the Hubble friction becomes comparable to the axion mass, the axion begins to roll, and its subsequent oscillations around the minimum of the potential are characterized by an energy density that is redshifted in a matter-like manner. This behavior is expected to persist until the present moment, thereby providing a natural mechanism for dark matter. 

This naive picture of the vacuum misalignment mechanism may be altered if the axion $a$ couples to a non-abelian gauge sector in a thermal bath via the operator, 
\begin{align}
{\cal L}\supset \frac{\alpha}{8\pi} \frac{a}{f_a} F^b_{\mu\nu}\widetilde{F}^{b\mu\nu}\,,
\label{Eq: Lagrangian first}
\end{align}
where  $F^b_{\mu\nu}$ is a gauge field strength tensor and $f_a$ the axion decay constant. Indeed, in the seminal work in Ref.~\cite{McLerran:1990de}, the existence of non-perturbative transitions in the high-temperature QCD plasma was demonstrated and their impact on the axion evolution was studied. These transitions describe thermal fluctuations in the topological charge that are reminiscent of the sphaleron processes in the electroweak sector. Despite some differences at the technical level, we will follow standard practice and refer to these thermal fluctuations as \textit{strong sphalerons} for simplicity. The main effect of the strong sphalerons consists in the appearance of an extra friction term $\Upsilon(T)$ in the axion equation of motion (EOM),
\begin{align}
{\ddot{a}}+\left[3 H + \Upsilon(T)\right]{\dot{a}}+
V'(a)=0\,,
\end{align}
where $H$ is the Hubble parameter. In the context of QCD, the authors of Ref.~\cite{McLerran:1990de} were able to show that the net effect of this new friction term turns out to be very weak.
It ends up being suppressed by small fermion Yukawa couplings in the Standard Model (SM), and thus the friction term is only active at high temperatures when the axion field is still frozen, leaving no significant impact on the prediction for the QCD axion relic density.

Nonetheless, thermal sphaleron transitions may play an important role in the axion evolution if they arise from a dark/hidden thermal bath, different from that of QCD and not containing any light fermions. The main aim of the present work is precisely to study in a general context how the axion dark-matter relic density may be modified due to thermal friction arising from such a dark sector. As we will show, in most cases,  sphalerons in a hidden sector result in a damping of the coherent axion oscillations and thus in a suppression of the axion abundance; still in some other scenarios, the friction can delay the onset of oscillations and thus enhance the axion abundance. We shall in particular derive an analytical formula for the adiabatic invariant that remains constant in the presence of thermal friction\,---\,the \emph{frictional adiabatic invariant}\,---\,taking into account the one-loop running of the gauge coupling. This will allow us to apply our novel mechanism\,---\,the \textit{frictional misalignmemt mechanism}\,---\,to a broad range of scenarios.

\definecolor{blueUnder}{HTML}{4169e1}
\definecolor{blueOver}{HTML}{20b2aa}
\begin{figure*}[ht]
    \centering
    \includegraphics[width=0.65\textwidth]{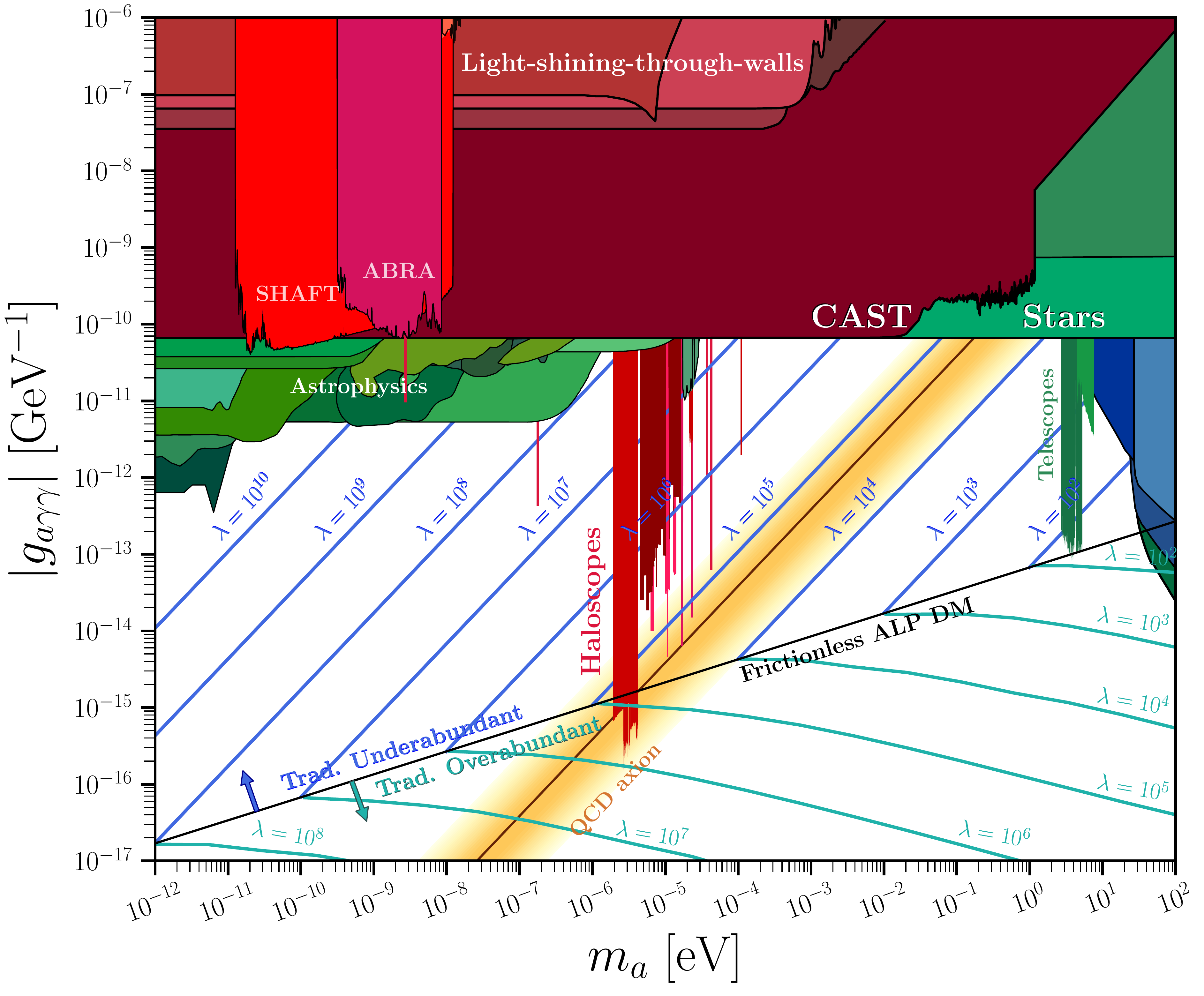}
    \caption{Axion coupling to photons $g_{a\gamma\gamma}\sim {\alpha}/({2\pi f_a})$ versus its mass for the case where the axion couples to two separate non-abelian gauge groups as studied in \cref{Sec:ALP2GaugeGroups}. The presence of friction can open up the ALP-dark-matter parameter space: For different values of the enhancement parameter $\lambda$, the correct ALP DM relic density is obtained for the traditionally underabundant region along the {\color{blueUnder} \bf blue} lines and for the traditionally overabundant region along the {\color{blueOver} \bf cyan} lines (assuming $\alpha_{\rm thr}=0.1$ and $g_{\rho,s}\sim {\cal O}(10)$, see  \cref{Sec:ALP2GaugeGroups} for details). Experimental bounds adapted from  \href{https://cajohare.github.io/AxionLimits/}{\color{black} \texttt{AxionLimits}}~\cite{AxionLimits}.}
    \label{fig:ALP-lambda}
\end{figure*}

Our work builds upon earlier work on thermal friction effects in cosmological axion models. In the past, such effects have been largely explored~\cite{Moore:2010jd,Laine:2016hma,Altenkort:2020axj} in other contexts, such as warm inflaton \cite{Berera:1995ie,Berera:1995wh,Berera:1999ws,Berera:1998px,Berera:2008ar,Bastero-Gil:2016qru,Berghaus:2019whh,Kamali:2021ugx,Berera:2020dvn,Yokoyama:1998ju,Visinelli:2011jy,Kamali:2019ppi,Mirbabayi:2022cbt}, late-time quintessence \cite{Berghaus:2020ekh}, leptogenesis \cite{Buchmuller:2005eh,Domcke:2020kcp}, electroweak baryogenesis \cite{Im:2021xoy} and early dark energy \cite{Berghaus:2019cls,Berghaus:2020ekh,Berghaus:2022cwf}; see Section~VI in Ref.~\cite{Agrawal:2022yvu} for a recent review. The present work also contributes to the larger effort in the community to explore the different possibilities for the cosmological axion evolution which open up once the assumptions of the canonical misalignment picture are relaxed or modified. Indeed, a variety of such scenarios have been proposed in the literature, e.\,g., large~\cite{Co:2018mho,Takahashi:2019pqf,Arvanitaki:2019rax,Huang:2020etx} or small~\cite{Dvali:1995ce,Banks:1996ea,Choi:1996fs,Co:2018phi} misalignment angles, parametric resonance~\cite{Co:2017mop,Harigaya:2019qnl,Co:2020dya}, kinetic misalignment mechanism~\cite{Co:2019jts,Chang:2019tvx,Co:2019wyp,Domcke:2020kcp,Co:2020jtv,Harigaya:2021txz,Chakraborty:2021fkp,Kawamura:2021xpu,Co:2021qgl,Co:2021lkc,Gouttenoire:2021wzu,Gouttenoire:2021jhk}, trapped misalignment~\cite{DiLuzio:2021gos}, axion fragmentation~\cite{Fonseca:2019ypl,Eroncel:2022abd,Eroncel:2022abc}, varying axion decay constant~\cite{Allali:2022yvx}, interaction with monopoles~\cite{Fischler:1983sc,Nakagawa:2020zjr} or modifications in the cosmological history of the Universe~\cite{Visinelli:2009kt,Arias:2021rer}, including entropy injection~\cite{Dine:1982ah,Steinhardt:1983ia,Choi:2022btl} or non-standard inflation scenarios~\cite{Dimopoulos:1988pw,Davoudiasl:2015vba,Hoof:2017ibo,Graham:2018jyp,Takahashi:2018tdu,Kitajima:2019ibn}.

The rest of the paper is organized as follows. In \cref{Sec:TheoreticalFramework}, we will provide the necessary foundation by reviewing the standard axion misalignment mechanism. We will also explore the constraints that apply to a cosmological hidden thermal bath. In \cref{Sec:FrictionalMis}, we will then analyze the consequences of the existence of a hidden thermal bath in the context of the vacuum misalignment mechanism. In \cref{Sec:MinimalALP}, we will apply the machinery developed in the second section to the most minimal ALP dark matter model, in which the hidden strong dynamics that generate the axion mass via instanton effects also yield the thermal friction. In \cref{Sec:ALP2GaugeGroups}, we will generalize these results by assuming that the gauge group that provides the friction is distinct from the one that provides the axion mass. \cref{Sec:QCDAxion}, finally, is dedicated to applying our results to the special case of the QCD axion. \cref{sec:conclusions} contains our conclusions and a comparison to other results in the literature.

\section{Framework and Assumptions}
\label{Sec:TheoreticalFramework}
\noindent\textbf{Vacuum misalignment mechanism\,---\,}%
%
Let us briefly review the prediction for the axion relic density in terms of the misalignment mechanism in the absence of any extra sources of friction.
In general, there are also other non-thermal production mechanisms for the axion such as the decay of topological effects. In this paper, we will, however, assume the pre-inflationary scenario and therefore focus solely on the misalignment mechanism.

The EOM for a classical, non-relativistic and homogeneous scalar field 
$\theta_a\equiv a/f_a$ in an expanding Friedmann-Lema\^itre-Robertson-Walker Universe reads 
\begin{align}
\ddot{\theta}_a+3 H \dot{\theta}_a+\frac{1}{f_a^2}V'(\theta_a)=0\,,
\end{align}
where $V'(\theta_a)={dV(\theta_a)}/{d\theta_a}$ and the spatial gradients have been neglected. 
For a field oscillating near its minimum, $V'(\theta_a)\simeq  m_a^2(T)f_a^2\, \theta_a$ is a good approximation, and the differential equation resembles that of a damped harmonic oscillator whose solution depends on the interplay between the friction term\,---\,the Hubble parameter $H(T)$\,---\,and the oscillator frequency\,---\,the axion mass $m_a(T)$. At high temperatures $T\gg \sqrt{m_a M_p}$, the axion field is frozen at an arbitrary initial misalignment angle $\theta_i$ due to Hubble friction, since $H(T)\gg m_a(T)$.  As the Universe cools down and the Hubble parameter decreases, the axion mass overcomes the friction and the field starts to oscillate at a temperature $T_{\rm osc}$ defined as $ H(T_{\rm osc})\sim m_a(T_{\rm osc})\equiv m_{\rm osc}$ where the exact numerical prefactors depend on the temperature scaling of the mass.

Using the Wentzel–Kramers–Brillouin (WKB) approximation, the final relic density can be expressed as
\begin{align}
  \frac{\rho_{a,0}}{\rho_{\rm DM}} \simeq 28  \sqrt{\frac{m_{a}}{\mathrm{eV}}}\sqrt{\frac{m_a}{m_{\rm osc}}} \left(\frac{\theta_{i}\,f_a}{ 10^{12}\, \mathrm{GeV}}\right)^{2} \mathcal{F}(T_{\rm osc})\,,
  \label{Eq:simple ALP relic dansity ratio CMB}
  \end{align}
  where $m_a\equiv m_a(T=0)$ is the axion mass at zero temperature,  $\mathcal{F}(T_{\rm osc}) \equiv\left(g_{\epsilon}(T_{\rm osc}) / 3.38\right)^{3/4}\left(3.93/g_{s}(T_{\rm osc})\right)$ is an $\mathcal{O}(1)$ factor, and $\rho_{\rm DM}\simeq 1.26\, {\mathrm{keV}}/{\mathrm{cm}^{3}}$ from Planck 2018 data~\cite{Aghanim:2018eyx}. 
  Through the factor $\sqrt{m_a/m_{\rm osc}}$, the relic density is affected by the temperature dependence of the axion mass, which may be constant $m_a(T)=m_a$, such that $\sqrt{m_a/m_{\rm osc}}=1$, or present a power-like dependence if the axion obtains its mass from a confining gauge group,
   \begin{align}
m_a(T)\simeq \left\{\begin{array}{lll} \displaystyle
m_a &\qquad \text{ for } \ \ \ T<T_{c}\\
\displaystyle 
\displaystyle m_{a}\,\left(\frac{ T_{c}}{T}\right)^{\beta} 
    & \qquad \text{ for } \quad T>T_c \,,
\end{array}\right.
\label{Eq: axion mass temp}
\end{align}
which leads to
\begin{align}
\sqrt{\frac{m_a}{m_{\rm osc}}}\simeq \left(\frac{\sqrt{m_a M_p}}{T_c}\right)^{\frac{\beta}{\beta+2}} \,.
\end{align}

\medskip
\noindent\textbf{Hidden thermal bath\,---\,}%
%
We shall assume the existence of a dark thermal bath characterized by a temperature $T'$ and composed of non-abelian $SU(N)$ gauge bosons in the absence of fermions. This hidden thermal bath is secluded from the SM one, and we remain agnostic as to the means of its generation. Eventually, the dark gauge group either confines or becomes spontaneously broken. We will explore both options in this work and analyze the possible outcomes.

In the case of confinement of a pure non-abelian gauge field, one generally expects the energy of the dark sector to be converted into glueballs. The glueballs subsequently evolve as dark matter and may in principle overclose the universe prematurely~\cite{Soni:2016gzf,Boddy:2014yra}. In order to avoid this, it is necessary to assume the decay of the glueballs to some light degrees of freedom such as moduli \cite{Halverson:2016nfq}. On the other hand, in the case of spontaneous symmetry breaking, the massive degrees of freedom are assumed to decay rapidly to a remaining unbroken U(1) subgroup of the initial $SU(N)$ gauge group. In either case, we assume that this process takes place rapidly so that from temperatures greater than the electroweak scale all the way down to recombination the energy density of the dark sector redshifts like radiation.

The temperature of this leftover radiation is constrained by limits on the effective number of neutrino species that can be inferred from the cosmic microwave background (CMB),
\begin{equation}
    \Delta N_{\rm eff}= \frac{8}{7}\left(\frac{11}{4}\right)^\frac{4}{3}\frac{\rho_X}{\rho_\gamma}\Bigg|_{T=T_{\rm rec}} < 0.3 \text{ at } 95\% \text{C.L.} \,,
    \label{eq:CMB-bound}
\end{equation}
where $\rho_\gamma$ is the energy of photons and $\rho_X$ is the energy of the byproducts of the dark gauge field and the bound comes from the TT,\,TE,\,EE,\,lowE\,$+$\,lensing\,$+$\,BAO Planck 2018 data~\cite{Aghanim:2018eyx}. We use $\rho^{(\prime)} = \pi^2/30 \,g_{\rho}(T^{(\prime)})\, T^{(\prime)4}$ for the energy density and $s^{(\prime)}= 2\pi^2/45\,g_{s}(T^{(\prime)})\,T^{(\prime)3}$ for the entropy density of each of the thermal baths. Here and in the rest of the paper, primed variables refer to the dark thermal bath.

Assuming that the entropy of the SM and the entropy of the dark sector are separately conserved (since they are not interacting with each other), one can relate the temperature ratio at recombination to the temperature ratio at some high reference temperature, $\xi \equiv T'_0/T_0$.
Under these assumptions, we find that
\begin{equation}
    \Delta N_{\rm eff}\simeq 4.4\frac{g'_{\rm \rho}(T_{\rm rec})}{g_{\rm \rho,SM}(T_{\rm rec})}\left[\frac{g'_{\rm s}(T_0)g_{\rm s,SM}(T_{\rm rec})}{g'_{\rm s}(T_{\rm rec})g_{\rm s,SM}(T_0)}\right]^{4/3}\,\xi^4\,,
\end{equation}
with $g_{\rm s, SM}(T_{\rm rec})= 3.91$, $g_{\rm \rho, SM }(T_{\rm rec})=3.36$, $g_{\rm s,SM}(T_0)=106.75$ and assuming $g'_{\rm s}=g'_{\rm \rho}$, we thus obtain
\begin{equation}
    \Delta N_{\rm eff}=0.016 \times n^{-1/3}\,\left(2N_c^2-2\right)^{4/3}\xi^4\,,
\end{equation}
where $n$ is the number of degrees of freedom of the byproducts and $N_c$ is the number of colors of the gauge field. With the simplest assumption, $n=2$, we can identify the dark thermal bath temperature as the SM temperature only for the $N_c=2$ case, while the $N_c=3$ case requires a small suppression of $T'_{0}$ compared to $T_{0}$ by about ten percent in order to be consistent with the upper limit set by CMB observations. Larger gauge groups would require a further suppression of the temperature ratio. This limit is also expected to be improved by CMB Stage-4 observations with a projected sensitivity of $\Delta N_{\rm eff}< 0.03$ \cite{CMB-S4:2016ple}. For concreteness, we will display our results for $N_c=3$ in the following and work with $\xi = 0.86$ which is the value that saturates the bound in \cref{eq:CMB-bound}. Note that, depending on the strength of the coupling between the axion and dark sector, it is possible that the axion may thermalize with the dark gauge field. In this case, one would need to add one degree of freedom for the axion in the dark thermal bath. This, however, only marginally changes the bound on the temperature ratio and hence, for simplicity, we disregard the thermalization of the axion with the dark sector. At lower temperatures, the relation between the temperature of the dark thermal bath and the temperature of the SM is well described by
\begin{align}
T'=\xi\left(\frac{g_{\rm s, SM}(T)\;g'_{\rm s}(T'_0)}{g_{\rm s, SM}(T_0)\;g'_{\rm s}(T')}\right)^{1/3} T \,,
\label{eq:temperature-relation}
\end{align}
where $g_{\rm s,SM}(T)$ describes the evolution of the entropic degrees of freedom of the SM, while the function $g'_{\rm s}(T')$ is a step function equal to $2(N_c^2-1)$ at temperatures higher than the confinement scale or the temperature of spontaneous symmetry breaking (whichever comes first) and equal to 2 at lower temperatures.

\section{Frictional misalignment}
\label{Sec:FrictionalMis}
We are now equipped to study the effect of friction due to the dark thermal bath on the axion evolution. The EOMs for the axion-gauge field system take the form,
\begin{align}
\ddot{\theta}_a+\left[3 H + \Upsilon(T')\right]\dot{\theta}_a&=-\frac{1}{f_a^2}V'(\theta_a) \,,\label{eq:eom}\\
\dot{\rho}_{\rm dr}+4H \rho_{\rm dr}&=f_a^2\,\Upsilon(T')\, \dot{\theta_a}^2 \,.\,\label{eq:eom2}
\end{align}
At weak gauge coupling, $\alpha \lesssim 0.1$, and if the Hubble rate is small compared to the rate of thermalization,
$H< \alpha^2 T'$, the sphaleron transitions induce an effective friction $\Upsilon(T')$ in the axion EOM that depends on the sphaleron rate $\Gamma_{\rm sph}$, whose general expression can be found in \cref{App:SphlaeronValerie}. For our purposes, the friction term is well approximated by
\begin{equation}
    \Upsilon(T')=\frac{\Gamma_{\rm sph}}{2 T' f_a^2}\simeq1.8\times\frac{N_c^2-1}{N_c^2}\frac{\left(N_c \alpha\right)^5 T'^3}{2  f_a^2}\,.
    \label{eq:friction}
\end{equation}
%

The axion potential will be assumed to be $V(\theta_a)=m_a^2(T) f_a^2 \left[1-\cos(\theta_a)\right]$, with temperature-dependent axion mass as in \cref{Eq: axion mass temp}. The value of the power-law coefficient $\beta$ depends on the specific theory: it is $\beta=4$ for the QCD axion in the dilute instanton gas approximation (DIGA) while for a pure $SU(N_c)$ gauge field it is given by $\beta=\frac{1}{2}\left(\frac{11}{3}N_c-4\right)$~\cite{Gross:1980br,Borsanyi:2015cka}.\footnote{ Note that, while the power-law behaviour predicted by the DIGA for QCD $\beta\sim 4$ agrees with some lattice simulations \cite{Borsanyi:2016ksw}, alternative approximations like the interacting instanton liquid model (IILM) suggest $\beta \sim 3.3$ and other lattice computations result in smaller values, $\beta \sim 1$ \cite{Trunin:2015yda} for QCD.}

For the application at hand, the energy of the gauge field is always much greater than the energy of the axion and hence the axion-induced backreaction in \cref{eq:eom2} will always be negligible. In practice, we therefore neglect \cref{eq:eom2} and only solve \cref{eq:eom} in the presence of a dark plasma that redshifts like radiation, $\rho_{\rm dr}\propto a^{-4}$ (where $a$ is the scale factor of the Universe). This assumption always holds in our scenario, as we checked a posteriori.

\medskip
\noindent\textbf{Running gauge coupling constant\,---\,}%
%
The dynamics in our mechanism span a substantial range of energies and therefore the running of the dark gauge coupling constant cannot be neglected. The coupling of the dark sector as a function of temperature may be approximated at one loop by
\begin{align}
\alpha\left(T'\right)=\frac{4\pi}{\bar{b}_0 N_c}\,\frac{1}{\ln\left(T'^2/\Lambda^2\right)} \,,
\label{eq:running}
\end{align}
where $\Lambda$ represents the confinement scale in the strongly coupled case and the \textit{would-be} confinement scale if the gauge group becomes spontaneously broken at energies above $\Lambda$. The factor $\bar{b}_0$ is related to the one-loop $\beta$-function
coefficient $b_0$ as $\bar{b}_0\equiv 4\pi b_0/N_c$ 
and takes the value $\bar{b}_0 = 11/3$ in the confining case, which only contains gauge bosons. On the other hand, in the case of spontaneous symmetry breaking, we assume the minimal Higgs content that allows us to break the $SU(N_c)$ down to $U(1)$. As outlined in \cite{Buccella:1979sk} this can be achieved with $N_c-2$ complex Higgses in the fundamental representation and one Higgs in the real adjoint representation of $SU(N_c)$. In that case, in the large $N_c$ limit, the beta function coefficient takes the value $\bar{b}_0 = 10/3$.

\medskip
\noindent\textbf{Onset of oscillations under thermal friction\,---\,}%
%
The first effect of the introduction of friction that we will study consists of a delay of the onset of oscillations. If at early times the friction is dominant, $\Upsilon(T'),\,3H\gg m_a(T)$, the axion field is frozen at its initial field value and the early time solution can be found by the "slow-roll" approximation which consists of neglecting the second derivative of the axion field in \cref{eq:friction}. The motion of the axion is then well described by \cite{Berghaus:2019cls}
\begin{align}
\theta_a(T)\simeq \theta_i {\rm e}^{-\frac{m_a(T_{\rm})^2}{(5+2\beta)\Upsilon(T')H(T)}}\,,
\end{align}
in the case where $\Upsilon(T')\gg 3H$, whereas it takes the form 
\begin{align}
\theta_a(T)\simeq \theta_i {\rm e}^{-\frac{m_a(T)^2}{6\left(2+\beta\right) H(T)^2}}\,,
\end{align}
when $3H\gg \Upsilon(T')$, where we approximate $g_s,\,g_\rho,\alpha\simeq {\rm const}$ and $\beta$ is defined in \cref{Eq: axion mass temp}.
The expression above indicates that the onset of oscillations takes place when both exponents are ${\cal O}(1)$. The precise value is best found by comparing with the numerical solution and identifying the prefactor that yields the most accurate results.

Using the above, we may write the condition for the onset of oscillations in the presence of friction as

\begin{align}
m_a(T_{\rm osc})\simeq \left\{\begin{array}{lll} \displaystyle
4 \,H(T_{\rm osc}) &\;,\; 3 H> \Upsilon \\
\displaystyle 
\displaystyle \frac{10 \Upsilon(T'_{\rm osc})\,H(T_{\rm osc})}{m_a(T_{\rm osc})} 
    & \;,\; 3 H < \Upsilon
\end{array}\right.\,,
\label{eq:oscillation-temperature}
\end{align}
depending on whether the thermal friction dominates over the Hubble friction at the onset of oscillations or vice versa. {Fundamentally, the onset of oscillations as defined in \cref{eq:oscillation-temperature} denotes the moment after which the energy density of the axion can no longer be approximated by a constant initial value given in terms of the misalignment angle and the axion parameters. After the onset of rolling the energy density varies appreciably}. The numerical prefactors on the right-hand side correspond to the values that yield the best agreement with the numerics regarding the late-time comoving number of axions and assume a QCD-like temperature dependence of the mass ($\beta=4$ in \cref{Eq: axion mass temp}). For other values of $\beta$, the prefactors are modified by ${\cal O}(1)$ factors.

For simplicity, in what follows we will stick to the assumption $\beta=4$ unless explicitly stated otherwise. While mild variations with respect to $\beta=4$ would lead to analogous conclusions, the exact dependence of the results on $\beta$, for values that deviate significantly from $\sim 4$, would require a more extensive analysis that is beyond the scope of this work; however, the present work provides all the ingredients necessary to facilitate such an extensive numerical study.

\medskip
\noindent\textbf{Frictional adiabatic invariant\,---\,}%
%
The second and most relevant effect of thermal friction on the axion evolution is the damping of the axionic oscillations, which results in a depletion of its relic abundance. In order to estimate the impact of this effect, we will now derive the quantity that remains constant during the axion evolution after the onset of oscillations — \emph{the frictional adiabatic invariant}.

For some general time-dependent friction $\Gamma(t)$, the EOM of the axion takes the form 
\begin{align}
\ddot{\theta}_a+\Gamma(t)\dot{\theta}_a+m_a^2(t)\theta_a=0\,,
\end{align}
where we expand the cosine of the potential to quadratic order. Via the change of variables
\begin{align}
\tilde{\theta}_a\equiv g(t)\,\theta_a\;\;\;,\;\;\;g(t) \equiv \;{\rm exp}\left[-\frac{1}{2}\int^t_{t_{\rm osc}}\Gamma(\tilde{t})d\tilde{t}\right]\,,
\end{align}
the EOM may be transformed into the standard template of a harmonic oscillator with a time-dependent frequency,
\begin{align}
\ddot{\tilde{\theta}}_a+\omega^2(t) \tilde{\theta}_a=0\,,
\end{align}
with $\omega^2=m_a(t)^2+\ddot{g}/g +\Gamma \,\dot{g}/g$. Provided then that $\dot{\omega}/\omega \ll \omega$, which will always be true in our case after the onset of oscillations as defined in \cref{eq:oscillation-temperature}, one may use the WKB approximation to write the solution at first order as
\begin{align}
\tilde{\theta}_a(t)\simeq\frac{\tilde{\theta}_0}{\sqrt{\omega}}\cos\left(\int^t_{t_{\rm osc}}\omega(\tilde{t})\;d \tilde{t}\right)\,,
\end{align}
and most importantly to obtain the adiabatic invariant,
\begin{align}
A & = \frac{\rho_\theta\left(t\right)}{\omega(t)}\,\exp\left[\int^td\tilde{t}\:\Gamma(\tilde{t})\right] = {\rm const} \,,
\end{align}
where $\rho_\theta$ is the energy density  $\rho_\theta/f_a^2 \equiv1/2 \dot{\theta}_{a}^{2} +1/2 m_{a}^{2}  \theta_{a}^{2}$.

If this general formula is applied to the standard case, in which the only friction is the Hubble expansion $\Gamma(t)=3H(t)$, then for $m_a\gg H$ and thus  $\omega\simeq m_a$, we obtain that the adiabatic invariant corresponds to the comoving number of axions $N_a$,
\begin{align}
N_a  = \frac{\rho_\theta\left(T\right)}{m_a(T)}\,\exp\left[\int^t_{t_{\rm osc}}d\tilde{t}\, 3 H(\tilde{t})\right] 
   = \frac{\rho_\theta a^3}{m_a} = {\rm const} .
\end{align}
In the presence of thermal friction $\Gamma=\Upsilon+3H$, and the adiabatic invariant  becomes

\begin{align}
\label{eq:adiabatic}
A_{\rm fr} 
  = \frac{\rho_\theta\left(T\right)a^3\left(T\right)}{m_a(T)} \exp\left[\int^t d\tilde{t}\:\Upsilon(\tilde{t})\right] = {\rm const} \,.
\end{align}

Notice that we are assuming that the frequency of oscillations can be approximated by $\omega\simeq m_a$. However, the frequency $\omega$ evaluated at the onset of rolling as defined in \cref{eq:oscillation-temperature} may deviate significantly from $m_a$. This generally introduces only an ${\cal O}(1)$ error in the final abundance as long as the thermal friction is not much greater than the Hubble friction by some ${\cal O}(10)$ value, evaluated at the onset of oscillations. This holds for the scenarios we consider in this work since  the onset of oscillations and the moment beyond which $\omega\sim m_a$ are in close proximity to each other due to the rapid decay of the ratio $\Upsilon(T')/m_a(T)\propto a^{-7}$ . For the cases in which there is a greater-than-${\cal O}(10)$ hierarchy between the thermal and Hubble friction, we will never use this formula to compute the abundance. In these cases of extraordinarily large friction at the onset of oscillations, the suppression of the axion abundance is so large that there is no motivation to study that parameter space for making the traditionally overabundant regime compatible with the observed dark matter and we can confidently say that within the parameter space of interest, the final abundance is zero. Instead, this limit will be relevant in the case of the traditionally underabundant regime; however, in that case the frictional adiabatic invariant will be irrelevant since we will assume the spontaneous breaking of the $SU(N_c)$ at some temperature greater than or equal to the onset of oscillations as defined in the lower branch in \cref{eq:oscillation-temperature}. The true onset of oscillation then occurs at the moment of spontaneous symmetry breaking instead.

The integral in the exponent $D\equiv \int d\tilde{t}\:\Upsilon$ can be computed analytically by plugging the expression for the thermal friction in \cref{eq:friction} and the one-loop running of the gauge coupling in \cref{eq:running}, as it is shown in \cref{sec:analytical_derivation_of_the_frictional_invariant}. The final results reads,

\begin{align}
&& D &= C \frac{M_p \Lambda}{f_a^2}\left[\frac{\tau^3 + \tau^2 + 2 \tau + 6}{\tau^4}\,e^\tau - \textrm{Ei}\left(\tau\right)\right]\;,\nn\\
&&  C &\simeq 10\frac{\pi^4}{\bar{b}_0^5}\left(\frac{g'_{\rm s}(T'_{0})}{g_{\rm s,SM}\left(T_0\right)g'_{\rm s}\left(T'_{\rm osc}\right)}\right)^{2/3}\frac{\sqrt{\bar{g}_{\rm \rho,SM}}}{\bar{g}_{\rm s,SM}^{1/3}}\,,\nonumber\\
&&\tau &= \ln\bigg(\frac{T'}{\Lambda}\bigg)\,,
\label{Eq:analytical integral}
\end{align}
where ${\rm Ei}(z)=-\int^\infty_{-z}dt\;{\rm e}^{-t}/t$ is the exponential integral function and $M_p$ denotes the reduced Planck mass. 

The derived \emph{frictional adiabatic invariant} in \cref{eq:adiabatic,Eq:analytical integral} remains constant from the onset of oscillations at $T'_{\rm osc}$ until the effects of friction are turned off at $T'_{\rm  end}$ due to either confinement or spontaneous symmetry breaking. Therefore, in order to compute the axion relic density, the result of the integral in \Cref{Eq:analytical integral} needs to be evaluated with the corresponding limits. This result is general and applies for any $SU(N_c)$ group and any value of $\beta$ in the temperature-dependent axion mass as long as the WKB approximation remains valid. The result can be collected and re-expressed in terms of physical scales as follows,
\begin{multline}
D \simeq 6.3 \left(\frac{10^8\;{\rm Gev}}{f_a}\right)^2\left(\frac{\Lambda}{150\;{\rm MeV}}\right)\\
\times \left[\frac{\tau^3 + \tau^2 + 2 \tau + 6}{\tau^4}\,e^\tau - \textrm{Ei}\left(\tau\right)\right]^{\tau_{\rm end}}_{\tau_{\rm osc}}\,,
\label{numerical-result}
\end{multline}
where the degrees of freedom were assumed to be approximately $g_{\rm \rho}\sim g_{\rm s}\sim {\cal O}(10)$, since our mechanism takes place mostly at high temperatures, and for concreteness, we selected $\bar{b}_0\sim 11/3$. Different options of these parameters will change the overall prefactor by some ${\cal O} (1)$ value.

If the running of the coupling constant is mild at the temperatures at which the friction is active, i.e. $T\gg \Lambda$, then the formula for the frictional invariant in \cref{Eq:analytical integral} can be further simplified to,
\begin{align}
    D &= 24\, C \frac{M_p \Lambda}{f_a^2}\left[\frac{T'/\Lambda}{\left[\ln(T'/\Lambda)\right]^5}\right]^{T'_{\rm end}}_{T'_{\rm osc}}\;\,.
\end{align}
as derived at the end of \cref{App:SphlaeronValerie}.

The set of \cref{eq:oscillation-temperature,Eq:analytical integral} constitute the main result of this paper and can be used to compute the axion dark-matter abundance for the various scenarios we will explore in the subsequent sections. The only case that is not covered by the formula above is the case in which the gauge group is spontaneously broken at some temperature before the onset of oscillations $T'_{\rm end}> T'_{\rm osc}$. In that case there is no suppression and we simply set $D=1$.

Our final result for the axion dark-matter abundance takes the form
\begin{align}
\frac{\rho_{a,0}}{\rho_{\rm DM}}\simeq 28 \sqrt{\frac{m_{a}}{{\rm eV}}} \sqrt{\frac{m_{a}}{m_{\rm osc}}}\left(\frac{\theta_i\;f_a}{10^{12}\,{\rm GeV} }\right)^2 {\rm e}^{-D} \left(\frac{m_{\rm osc}}{4\,H_{\rm osc}}\right)^{3/2} {\cal F}\nonumber\\
\label{eq:relic-abundance}
\end{align}

The various factors have been rearranged so that the result matches the standard result in \cref{Eq:simple ALP relic dansity ratio CMB} in the absence of the factors ${\rm e}^{-D}$ and $\left(\frac{m_{\rm osc}}{ 4 H_{\rm osc}}\right)^{3/2}$. These factors correspond respectively to an overall suppression due to friction and an enhancement due to the delay in the oscillations in the overall abundance. It will be shown in the various realizations of our mechanism that either of these factors may dominate depending on the scenario. The rest of the paper is devoted to applying the main results in \cref{eq:temperature-relation,eq:friction,eq:running,eq:oscillation-temperature,eq:relic-abundance} for the computation of the relic density in various scenarios.

\begin{figure}
    \centering
    \includegraphics[scale=0.4]{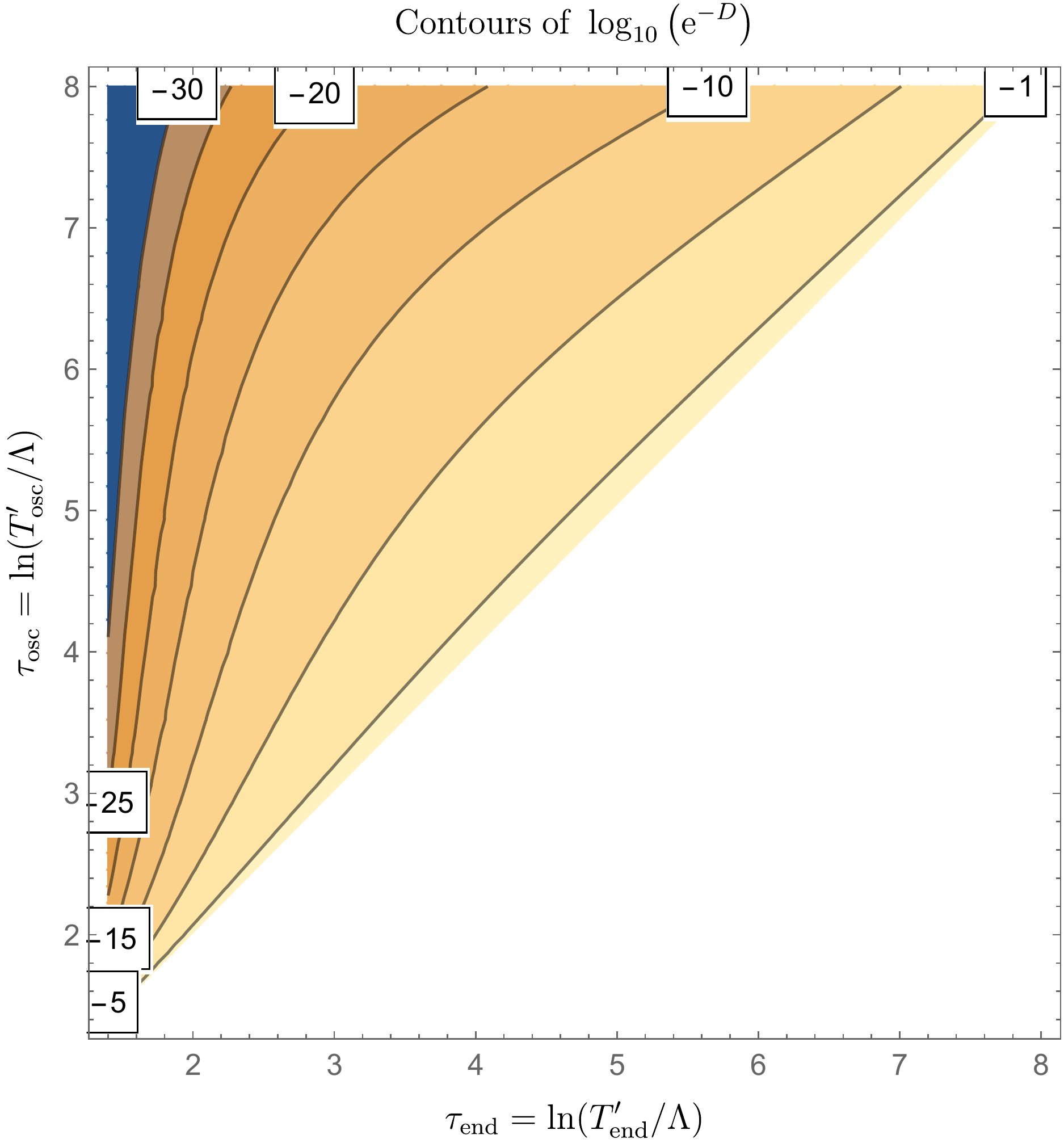}
    \caption{Contour plot of the overall suppression factor ${\rm e}^{-D}$ where $D$ given by \cref{numerical-result} for $f=10^8\;{\rm GeV}$ and $\Lambda=150\; {\rm MeV}$.}
    \label{fig:exponent}
\end{figure}
\section{Minimal ALP Dark Matter Model}
\label{Sec:MinimalALP}

The first scenario we will focus on is the most minimal ALP dark-matter model. In this scenario, there is a single non-abelian gauge group that provides a mass to the ALP via the anomaly while simultaneously introducing the thermal friction term in the axion EOM described in the preceding section. We will investigate its impact on the axion relic abundance. 
The relevant Lagrangian corresponds to that in \cref{Eq: Lagrangian first} and the temperature-dependent mass generated by this anomalous coupling corresponds to that in \cref{Eq: axion mass temp}, where the critical temperature corresponds to the confinement scale of the gauge group $T_c=\Lambda$ and the zero temperature mass of the axion is $m_a\simeq {\Lambda^2}/{f_a}$. For concreteness, in the numerical results, we will further assume that $\beta\sim 4$ and that the gauge group\footnote{For a pure $SU(3)$ gauge group DIGA predicts $\beta=7/2$, which does not modify significantly the numerical results.} is $SU(3).$ 

The final ingredient that needs to be specified in order to apply the machinery that was developed in the preceding section is the temperature $T'_{\rm end}$ at which the friction term turns off. For $T'\ll \Lambda$, the sphaleron rate is exponentially suppressed, but the expression in \cref{eq:friction} is no longer valid for these temperatures (the approximations break down for $\alpha (T)\gtrsim 0.1$). There is thus some ambiguity in regard to the value of the gauge coupling that signals the end of the effect of friction, $\alpha_{\rm thr}\equiv \alpha(T'_{\rm end})$. One may be tempted to be conservative and choose $\alpha_{\rm thr}=0.1$ so that the friction term is completely turned off whenever we are outside the validity of the formula for the sphaleron rate. For this case, thermal friction is only active far above the confinement scale, and even a large friction at such early times would not be imprinted in the final abundance, because at that point, the axion has not yet started to oscillate. Nevertheless, we argue that this conservative option underestimates the total effect. Indeed, we expect a non-zero sphaleron rate for larger values of the gauge coupling until $\alpha_{\rm thr}\sim 1/3$, even though the accuracy of the sphaleron rate formulas is lost. This approach is consistent with other attempts in the literature to extrapolate the sphaleron rate expression to couplings greater than $\alpha_{\rm thr}\sim 0.1$, such as in the case of heavy-ion collisions. Ref.~\cite{Kapusta:2020qdk} concluded that the sphaleron rate remains significative at least up to $\alpha_{\rm thr}\sim 1/3$, albeit the rate may be suppressed by up to an order of magnitude with respect to \cref{eq:friction}. For this reason, we compute the overall axion abundance for two distinct values of $\alpha_{\rm thr}=0.2,\,0.4$, thus demonstrating that the result is quite sensitive to this choice. 
  
\begin{figure}
    \centering
    \includegraphics[scale=0.30]{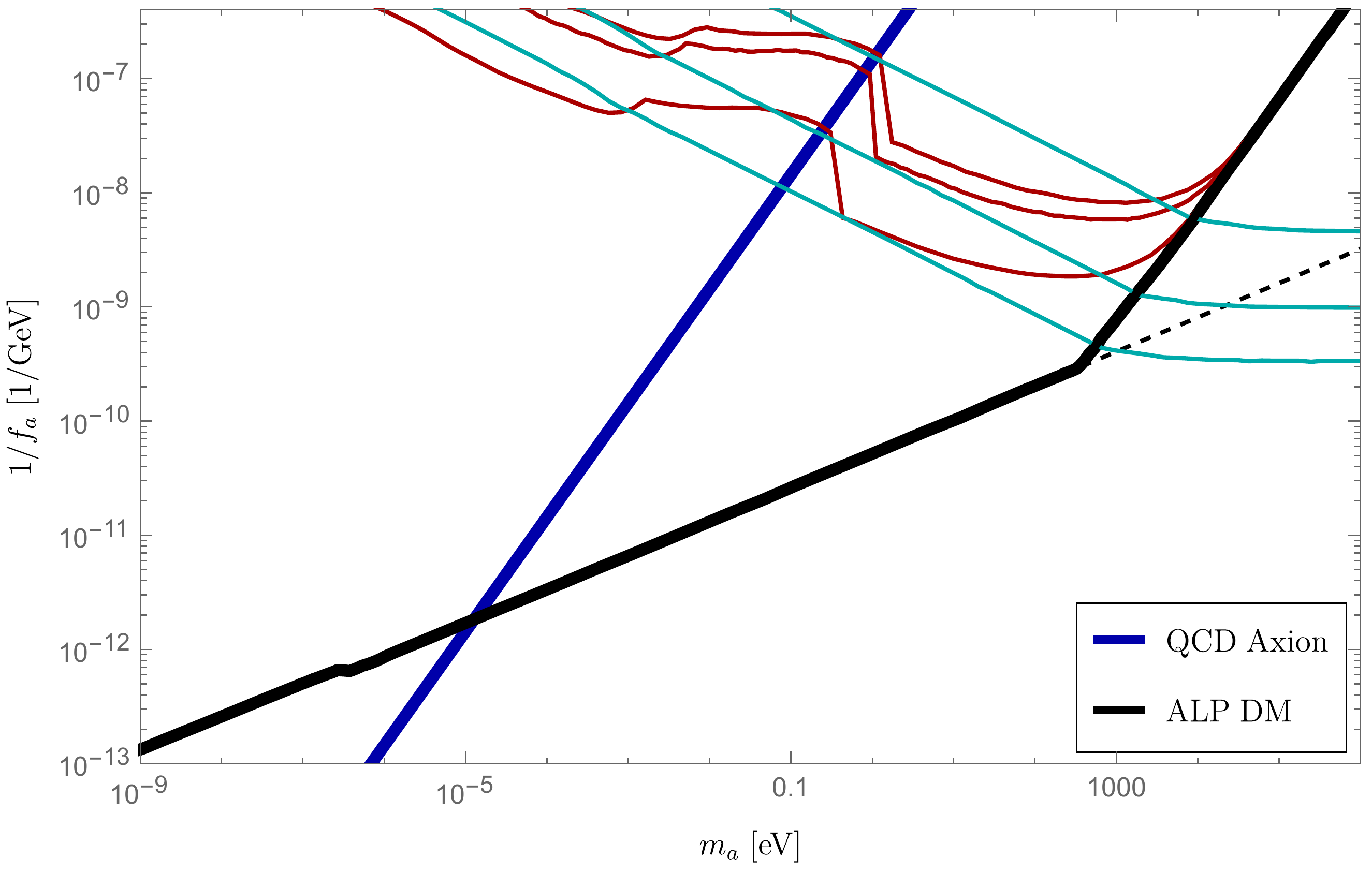}
    
    \includegraphics[scale=0.30]{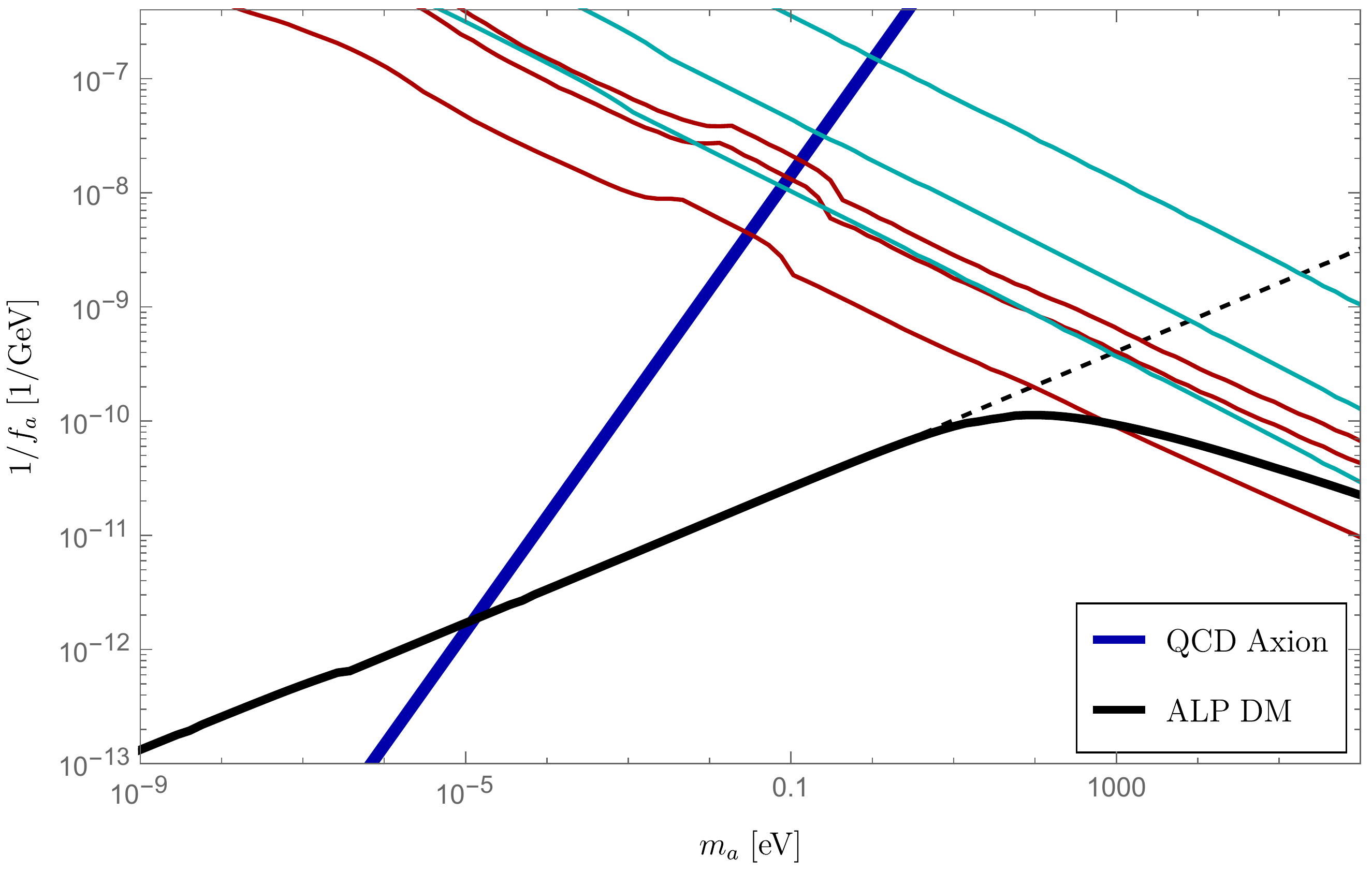}
    \caption{
    Predictions for the values of $\{m_a,\,1/f_a\}$ that reproduce the correct axion DM relic density (black solid lines), taking into account thermal friction for $\alpha_{\rm thr}=0.2$ (\textit{top panel}) and $\alpha_{\rm thr}=0.4$ (\textit{bottom panel}). They deviate with respect to the standard frictionless computation, which is shown in dashed black lines. The red lines are constant depletion factor ${\rm e}^{-D}$, with $D$ defined in \cref{Eq:analytical integral} and correspond to $10^{-20}$, $10^{-10}$, $10^{-1}$ from top to bottom. The cyan lines are lines of constant enhancement factor $\left(m_{\rm osc}/4 H_{\rm osc}\right)^{3/2}$ and correspond to 100, 10, 2 respectively from top to bottom.}
    \label{fig:ALPDM}
\end{figure}
  
The results are shown in \cref{fig:ALPDM}, where the points in the $\{m_a,\,1/f_a\}$ plane that can successfully account for the observed dark matter are displayed. Interestingly, depending on the value of $\alpha_{\rm thr}$ the resulting relic density is either suppressed or enhanced with respect to the frictionless case. For $\alpha_{\rm thr}=0.2$, the dominant effect of the friction is the delay of the onset of oscillations that results in an enhancement of the relic density and thus requires smaller decay constants $f_a$ to reproduce the observed DM density (the black solid line in the top panel of \cref{fig:ALPDM} bends upwards). Instead, for $\alpha_{\rm thr}=0.4$, the dominant effect is the damping of the oscillations that reduces the relic density requiring larger values of $f_a$ (the black solid line in the bottom panel of \cref{fig:ALPDM} bends downwards).
  
\begin{figure}
    \centering
    \includegraphics[scale=0.45]{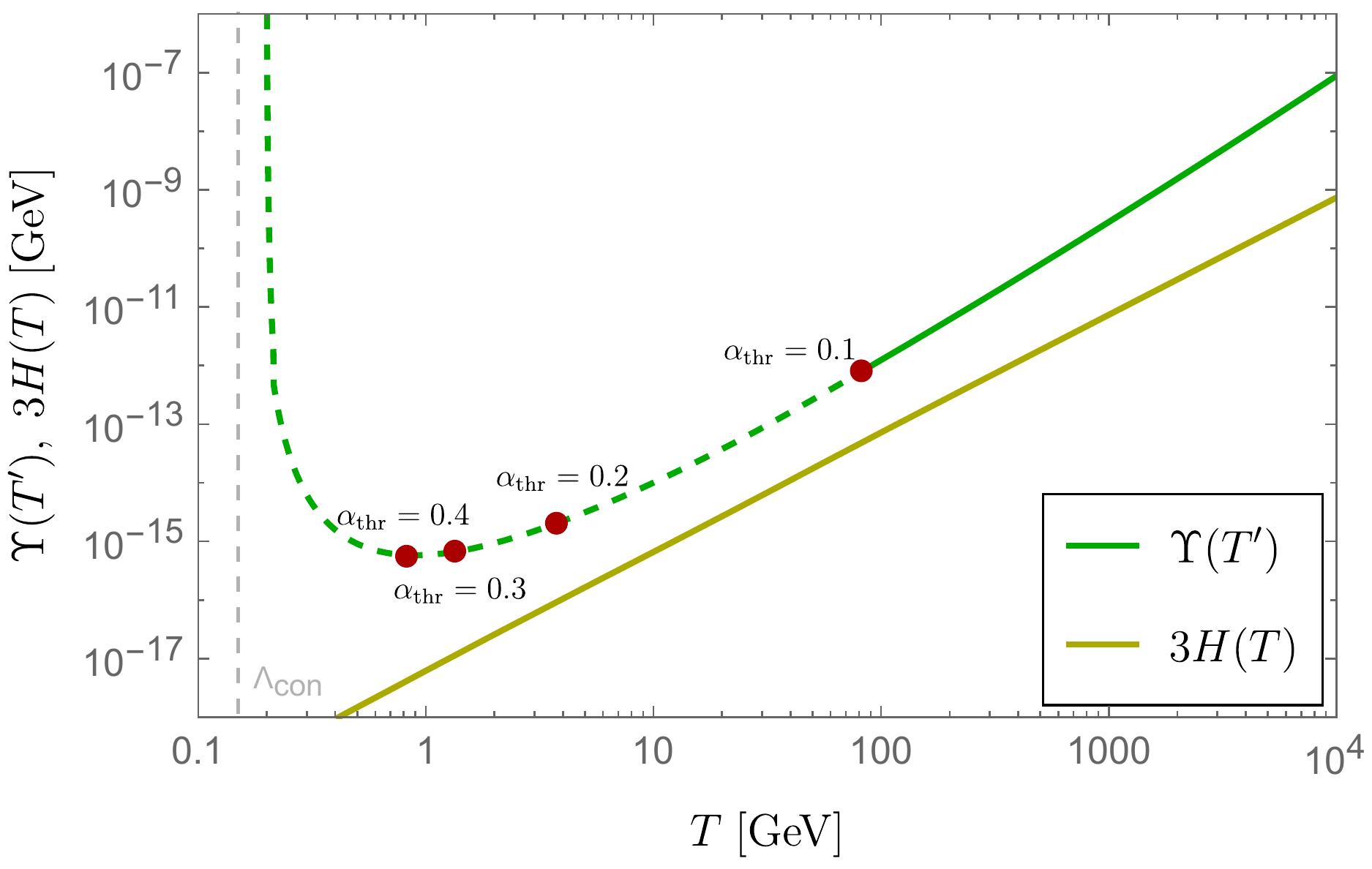}
    \caption{Gauge friction versus Hubble friction for some example parameter values. The red dots denote the instant we switch off the friction effects depending on the largest acceptable value of the gauge coupling $\alpha_{\rm thr}$.}
    \label{fig:YvsH}
\end{figure}

Regarding the validity of our results, even though we are extrapolating the sphaleron rate, it is important to note that suppressing it by an order of magnitude would only marginally affect our results. The important point here is that allowing for the extrapolation of the sphaleron rate to higher couplings allows us to capture the effect of friction for temperatures closer to the confinement scale, while the results are only mildly sensitive to the exact value of the friction coefficient. This exact effect is displayed in \cref{fig:YvsH}. One may observe that by extrapolating our expression for higher values of the gauge coupling, the effects of friction last for a much longer period, whereas if we switch off the friction at $\alpha_{\rm thr}=0.1$, the corresponding temperature is too premature for the friction effects to act on the axion after it has started to roll.
Of course the sphaleron rate (green line) is inaccurate at lower temperatures than the one corresponding to the point $\alpha_{\rm thr}=0.1$ but since one does not expect it to be exponentially suppressed yet, any ${\cal}O(10)$ suppression of the sphaleron rate would only imply a small modification of the lines in \cref{fig:ALPDM}.

The main conclusion from our analysis is that the predictions for the dark-matter abundance of the standard calculation in the Minimal ALP scenario are only reliable up to a mass of approximately $m_a\simeq 10^2\;{\rm eV}$ and for greater masses  the friction plays an important role and we find a deviation from the standard prediction. This conclusion holds true as long as the axion obtains its mass from instanton effects of a dark non-abelian gauge sector. 
Further studies on the extrapolation of the sphaleron rate close to the confinement scale are needed in order to elucidate the maximum value of the coupling constant at which the friction is still active, which strongly impacts the axion relic density prediction.

Finally, regarding the phenomenological consequences of this minimal model, it is important to note this minimal ALP cannot couple to photons for the parameter region where the friction terms are important; the reason being that for those values of $\{m_a,\,1/f_a\}$ and assuming an electromagnetic anomaly coefficient $E\sim \mathcal{O}(1)$ \cite{Georgi:1986df}, the corresponding value of the axion coupling to photons is excluded due to several cosmological considerations (see e.g. Ref.~\cite{Cadamuro:2011fd}).

\section{ALP coupled to two Gauge groups}
\label{Sec:ALP2GaugeGroups}

As another application of the thermal friction effects, we will consider the case in which the potential of the axion and the friction are provided by two separate gauge groups. We will allow for a possible coupling hierarchy among them that could arise in the context of clockwork axions or alignment scenarios \cite{Kim:2004rp,Choi:2014rja,Choi:2015fiu,Kaplan:2015fuy,Giudice:2016yja,Long:2018nsl}. The corresponding axion interaction Lagrangian reads,
\begin{align}
{\cal L_{\rm int}}= \frac{\alpha_G}{8\pi}\theta_a G^b_{\mu\nu}\widetilde{G}^{b\mu\nu} 
+ \lambda \frac{\alpha}{8\pi}\theta_a F^b_{\mu\nu}\widetilde{F}^{b\mu\nu}\,,
\label{Eq: Lagrangian first2}
\end{align}
where $G^b_{\mu\nu}$ is the field strength tensor of the gauge group that confines and provides the potential through instanton effects and $F^b_{\mu\nu}$ is the field strength tensor of the gauge group that provides the friction and will be assumed to be spontaneously broken. The spontaneous breaking of the gauge field $F^b_{\mu\nu}$ allows us to have large friction at early times, while large contributions to the potential are avoided at late times, since the instanton effects are exponentially suppressed after the gauge is spontaneously broken. We will also assume a hierarchy between the couplings, which is encoded in the parameter $\lambda\gtrsim1$. Such a value of the enhancement parameter may be justified by the alignment mechanism if the enhancement is relatively small \cite{Kim:2004rp}. However, we will also explore very large hierarchies, $\lambda \gg 1$, which is possible in the context of the clockwork mechanism \cite{Choi:2014rja,Kaplan:2015fuy,Giudice:2016yja,Long:2018nsl}, since it grows exponentially with the number $N$ of scalar fields of the full theory, $\lambda=3^N$. In principle, there is no bound on $N$, and hence we will also explore very large $\lambda$ values.

Upon confinement, the axion mass exhibits the temperature dependence in \cref{Eq: axion mass temp} for $T_c=\Lambda_G$ and $m_a\simeq{\Lambda_G^2}/{f_a}$, where $\Lambda_G$ is the confinement scale. Note that this scale differs from $\Lambda$, which corresponds to the would-be confinement scale of the gauge group that causes the friction and which becomes spontaneously broken, see \cref{eq:running}. Taking the above into account and assuming that the onset of oscillations corresponds to $m_{\rm osc}\equiv m_a(T_{osc})\simeq 4 H(T_{osc})$, the prediction for the relic density in  \cref{Eq:simple ALP relic dansity ratio CMB} for the frictionless case can be simplified to,
\begin{align}
\frac{\rho_{a,0}}{\rho_{\rm DM}} \simeq \left(\frac{f_a}{7.57 \cdot 10^{9}\; {\rm GeV}}\right)^{5/3}\,\theta_i^2\,\sqrt{\frac{m_a}{{\rm eV}}}\,\frac{\mathcal{F}(T_{\rm osc})}{g_{\rm \rho, SM}(T_{\rm osc})^{1/6}}\,.
\label{eq:ALPDM}
\end{align}
This implies that, for initial misalignments $\theta_i\sim\mathcal{O}(1)$, the ALP abundance correctly accounts for dark matter only for axion scales $f_a \sim 7\cdot 10^{9}\; {\rm GeV}\; ({\rm eV}/m_a)^{3/10}$, while it is underproduced (overproduced) for smaller (larger) $f_a$. The thermal friction effects studied in this work may in principle allow us to open up the parameter space for ALP dark matter. In the traditionaly underabundant case, this is possible if the onset of oscillations is delayed with respect to the standard case, while in the traditionally overabundant case, this may happen if the ALP experiences the gauge friction after it has started rolling. We will now investigate both possibilities separately.

\medskip
\noindent\textbf{Underabundant ALP\,---\,}%
%
In the presence of thermal friction, the true temperature at which the onset of oscillations occurs might be distinct from the standard one and correspond to the lower option in \cref{eq:oscillation-temperature}. Treating $T_{\rm osc}$ as a free parameter we may derive a new expression for the relic abundance,
\begin{align}
\frac{\rho_{\rm a,0}}{\rho_{\rm DM}}\simeq \left(\frac{m_a f_a}{T_{\rm osc}^2}\right)^4 \,\theta_i^2\,\left(\frac{T_{\rm osc}}{4.53\cdot10^{-10}\,{\rm GeV}}\right) \frac{{\cal F}}{g_{\rm \rho,SM}(T_{\rm osc})^{3/4}}\,.\nonumber\\
\label{eq:relic-abundance-ALP}
\end{align}
For $\theta_i\simeq 1$, it follows that the required temperature $T_{\rm crit}$ for the onset of oscillations that predicts the correct dark matter density reads
\begin{align}
T_{\rm crit}\simeq 21.6\,{\rm GeV}\left(\frac{m_a\,f_a}{{\rm GeV}^2}\right)^{4/7} \frac{{\cal F}^{1/7}}{g_{\rm \rho,SM}(T_{\rm crit})^{3/28}}\,.
\label{eq:critical-temperature}
\end{align}

In the presence of thermal friction, it is possible to delay the onset of oscillations until this critical temperature provided the friction is strong enough to prevent the axion from rolling until that moment. The relevant scales of the problem must then obey the following hierarchy,
\begin{align}
\Lambda\leq  T_2 \leq T_{\rm end} &=  T_{\rm crit} \leq T_1
\label{eq:temperature-conditionsALP}
\end{align}
where $T_1$ and $T_2$ are the solutions to the upper and lower cases in \cref{eq:oscillation-temperature}, respectively.

Our scenario assumes that spontaneous breaking occurs at just the right moment and from that moment on the ALP undergoes oscillations as usual. Aside from the aforementioned scenario, there is also a possibility that the scale of spontaneous symmetry breaking is lower than $T_2$. In that case the axion abundance first overshoots the desired one, since the axion remains frozen beyond temperature $T_{\rm crit}$, but that additional abundance is also diluted after the onset of oscillations. We exclude this fringe possibility because it always requires severe fine-tuning in the closeness of $T_2$ and $T_{\rm crit}$.

We can now use \cref{eq:oscillation-temperature} to write 
\begin{align}
T_2<T_{\rm crit} \;\;\longleftrightarrow \;\;\frac{10 \Upsilon(T'_{\rm crit})\,H(T_{\rm crit})}{m_{a}(T_{\rm crit})^2}>1\,,
\end{align}
which can be recast as 
\begin{align}
\frac{{\cal F}_a\left(\frac{ m_a f_a }{17.0 \,{\rm GeV}^2}\right)^{10/7} \lambda^2}{\left[1+0.17\left(\ln\left[{\cal F}_b\left(\frac{ m_a f_a}{{\rm GeV}^2}\right)^{1/7}\right]+\ln\left[{\Lambda_G^2}/{\Lambda^2}\right]\right)\right]^5} > 1\,,
\label{eq:enhancement-parameter}
\end{align}
where ${\cal F}_a$ and ${\cal F}_b$ depend on the effective number of degrees of freedom and are given by
\begin{align}
{\cal F}_a&=\frac{g_{\rm s,SM}(T_{\rm crit})\,g'_{\rm s}\left(T'_{0}\right){\cal F}^{13/7}}{g_{\rm s,SM}(T_0)\,g'_{\rm s}\left(T'_{\rm crit}\right)\,g_{\rm \rho,SM}(T_{\rm crit})^{25/28}}\,,\\
{\cal F}_b&=\left(\frac{g_{\rm s,SM}\left(T_{\rm crit}\right)\,g'_{\rm s}\left(T'_{0}\right)}{g_{\rm s,SM}(T_0)\,g'_{\rm s}\left(T'_{\rm crit}\right)}\right)^{2/3}\frac{{\cal F}^{2/7}}{g_{\rm \rho,SM}(T_{\rm crit})^{3/14}}\,.\nonumber
\end{align}
The expression in \cref{eq:enhancement-parameter} displays a necessary condition for the delay of the onset of oscillations to lead to the correct DM abundance. Aside from the small dependence on the degrees of freedom, this is an expression that depends on the enhancement parameter $\lambda$, the ratio of the confinement scales $\Lambda_G/ \Lambda$ and the product $m_af_a$.

It is important to note that this analysis has focused only on the axion zero mode as a first step. Due to the delay on the onset of oscillations, at $T_{\rm osc}$ the Hubble parameter is significantly smaller than the axion mass and thus the axion field experiences large amplitude oscillations before being significantly damped. As a consequence, the non-harmonicities of the potential will induce the production of higher momentum axion quanta, i.e. axion fragmentation \cite{Fonseca:2019ypl,Eroncel:2022abd}. This process is not expected to significantly modify the prediction for the axion relic density (and thus will not be discussed further in the present work) but could give rise to observational consequences pointing to a mechanism responsible for the delay of the onset of oscillations \cite{Eroncel:2022abc}.

The main result of this section is provided in \cref{fig:ALP-lambda} where we show that the region of the parameter space which traditionally results on axion underabundance, may now account for all the dark matter provided that the enhancement parameter $\lambda$ takes at least the indicated value (which is required to guarantee that the axion remains frozen until the critical temperature). For the purpose of producing this plot, we followed a conservative approach and assumed that our approximate expressions for the friction term are valid until $\alpha_{\rm thr}=0.1$. If we relax this assumption, one requires a smaller enhancement. At this point it is important to mention that the results shown above do not apply necessarily for the QCD axion which will be treated separately in the next section. We simply added the QCD axion line in figure \cref{fig:ALP-lambda} as a visual reference point.

To summarize, our analysis shows that in the presence of thermal friction it is possible to delay the onset of oscillations sufficiently enough to enhance the overall ALP relic abundance. This enhancement may be enough to account for all, or a fraction of dark matter and it is in principle possible to open up the entirety of the parameter space in which the ALP relic abundance from the misalignment mechanism is too small, provided that the coupling to the dark non-abelian gauge field is strong enough. 

\medskip
\noindent\textbf{Overabundant ALP\,---\,}%
%
In contrast to the previous case, in the traditionally overabundant case, one requires the axion to experience the effects of friction after the onset of oscillations. Exploring the parameter space for this scenario is a straightforward application of the general formulas derived in the previous chapter. It is however difficult to express the results in closed analytical form because of the complexity involved in solving the relevant transcendental equations. For clarity, we will simplify the equations of the previous chapter by approximating the degrees of freedom at $T_{\rm osc}$ by an ${\cal O}(10)$ number and assuming that $H_{\rm osc}$ corresponds to the lower branch of \cref{eq:oscillation-temperature}. The axion relic density in \cref{eq:ALPoverRHO} then translates into,
\begin{align}
\frac{\rho_{\rm a,0}}{\rho_{\rm DM}}&=0.11 \,{\rm e}^{-D}\left(\frac{m_a^{10}f_a^{10}\lambda^{14}}{{\rm GeV}^{20}\,\tau_{2}^{ 35}}\right)^{1/13}\,,\label{eq:ALPoverRHO}
\end{align}
where the exponential suppression parameter $D$ in \cref{numerical-result} now reads,
\begin{multline}
D \simeq 6.3 \left(\frac{\lambda\,10^8\;{\rm Gev}}{f_a}\right)^2\left(\frac{\Lambda}{150\;{\rm MeV}}\right)\\
\times \left[\frac{\tau^3 + \tau^2 + 2 \tau + 6}{\tau^4}\,e^\tau - \textrm{Ei}\left(\tau\right)\right]^{\tau_{\rm end}}_{\tau_{2}}\,,
\label{eq:numerical-result2}
\end{multline}
and the temperature of onset of oscillations is
\begin{align}
\tau_{2}\equiv \ln\left({T'_2}/{\Lambda}\right) &=\ln\left[\frac{13.2 \;{\rm GeV}}{\Lambda}\left(\frac{ m_a^6 f_a^6\,\tau_{2}^{ 5}}{{\rm GeV}^{12}\lambda^2}\right)^{1/13}\right]\label{eq:ToscY}\,.
\end{align}
For convenience, let us also define the variables $\tau_1$ and $\tau_{\rm thr}$ as
\begin{align}
\tau_{1}&\equiv \ln\left({T'_1}/{\Lambda}\right)=\ln\left(\frac{600\,{\rm GeV}^{1/6}\,f_a^{1/3}\,m_a^{1/2}}{\Lambda}\right)\label{eq:ToscH}\\
\tau_{\rm thr}&=\frac{\pi}{5\,\alpha_{\rm thr}}\label{eq:Tendalpha}
\end{align}
where  $\tau_1$ corresponds to the solution to the upper branch of \cref{eq:oscillation-temperature} and $\tau_{\rm thr}$ corresponds to the value of $\tau$ when the coupling constant reaches the threshold value $\alpha_{\rm thr}$. The required values of $\lambda$ and $\Lambda$ to yield the right dark-matter relic abundance can then be determined using the following algorithm.

\begin{itemize}
    \item We select a point of interest in the $\{m_a,f_a\}$ plane in the overabundant regime, $f_a > 7.57\cdot 10^{9}\; {\rm GeV}\; ({\rm eV}/m_a)^{3/10}$.
    \item We evaluate the first line of \cref{eq:numerical-result2} for the selected point and determine the range of values in the $\{\lambda,\Lambda\}$ plane that yield a value that is at most of order $\sim 10$. This is required in order to avoid fine-tuning of $\tau_{\rm end}$. 
    \item Next, for a choice of $\{\lambda,\Lambda\}$ we evaluate both \cref{eq:ToscY} and \cref{eq:ToscH} and demand that $\tau_{1}>\tau_{2}$ so that the gauge friction is comparable to Hubble friction at the onset of oscillations. We repeat the search in the $\{\lambda,\Lambda\}$ plane until an acceptable pair of values is found.
    \item Next, we set the left hand side of \cref{eq:ALPoverRHO} equal to one and solve for $\tau_{\rm end}$.
    \item Finally, if $\tau_{\rm end}\geq\tau_{\rm thr}$ then the values $\{\lambda,\Lambda\}$ chosen are acceptable and the correct dark-matter abundance is obtained. Otherwise we go back to step 3 and try a different choice of $\{\lambda,\Lambda\}$.
\end{itemize}

Following this algorithm we have produced \cref{table:1} where some choices of parameters which yield the correct dark-matter abundance are displayed.

\begin{table}[h!]
\centering
\begin{tabular}{|c| c| c| c| c| c| c|} 
 \hline
 No & $f_a$ (GeV) & $m_a$ (eV) & $\lambda$ & $\Lambda$ (GeV) & $\tau_{2}$ & $\tau_{\rm end}$ \\
 \hline
 1 & $10^{15}$ & 1 & $6.3\times10^{6}$ & 1 & 7.31 & 6.54 \\ 
 2 & $10^{14}$ & $10^{-3}$ & $4.2\times 10^{6}$ & $5\times 10^{-3}$ & 8.48 & 6.85 \\
 3 & $10^{17}$ & $10^{-11}$ & $2\times 10^{11}$ & $3\times10^{-6}$ & 8.95 & 7.82\\ 
 \hline
\end{tabular}
\caption{Sample values of $\{\lambda,\Lambda\}$ that may induce sufficient friction to deplete the axion relic abundance to the observed one for a given choice of $\{m_a,f_a\}$.}
\label{table:1}
\end{table}

One may proceed in an analogous manner to study any value in the $\{m_a,f_a\}$ plane and determine the coupling strength and confinement scale required to account for the observed dark-matter abundance. As one may expect from \cref{eq:numerical-result2} large values of $f_a$ require a large enhancement parameter $\lambda$. This fact is manifest in the values displayed in \cref{table:1}.

In order to get a more complete picture of the relevant parameter space, we may also find the lowest value of $\lambda$ that sufficiently dilutes the abundance to the observed value. The lowest possible value of the enhancement parameter corresponds to the minimum friction, acting on the axion for the maximum time period while still accounting for the observed dark-matter abundance. With this in mind we select the latest possible moment to turn off the friction by identifying $\tau_{\rm end}=\tau_{\rm thr}$ and we are also setting $\Lambda$ to be much lower than the confinement scale of the group that provides the mass
\begin{equation}
    \Lambda\ll \Lambda_G\sim\sqrt{m_a f_a}\,.\label{eq:condALP}
\end{equation}
As long as \cref{eq:condALP} is satisfied, the exact choice of $\Lambda$ changes the result for the enhancement parameter by some ${\cal O}(1)$. At this point we are mainly interested in the minimum order of magnitude for the enhancement parameter and hence, for concreteness, we select the value 
\begin{equation}
    \Lambda={\rm e}^{-10}\times 600\,{\rm GeV}^{1/6}\,m_a^{1/2}\,f_a^{1/3}\,,
\end{equation}
which automatically satisfies the condition in \cref{eq:condALP}  in the parameter space of interest. These choices considerably simplify the algorithm mentioned above and yield a unique solution for $\lambda$ which is roughly the minimum required to dilute the dark-matter abundance to the observed one. The parameter space is displayed in \cref{fig:ALP-lambda}.

Our results demonstrate that the traditionally overabundant region of ALP dark matter parameter space may be opened up in the presence of thermal friction. Unlike the case of the traditionally underabundant regime where the minimum value of the enhancement parameter depends on the product $m_a f_a$ (and thus the constant $\lambda$ lines in \cref{fig:ALP-lambda} are parallel to the QCD axion band), for the traditionally overabundant the required $\lambda$ strongly depends on the value of the axion decay constant $f_a$ and to a lesser degree on the axion mass.

\section{QCD Axion}
\label{Sec:QCDAxion}

In this section, we study the effect of thermal friction on the QCD axion and apply the results of the previous sections for this particular case. The QCD axion is particularly interesting due to its possible role in resolving the strong CP problem. Some of the results of the previous section are not directly applicable to this case since, for the canonical QCD axion, the mass has an extra suppression factor due to the existence of light quarks,
\begin{align}
m_a^{\rm QCD} f_a=m_\pi f_\pi \frac{\sqrt{z}}{1+z},\;\;{\rm where}\; z\equiv\frac{m_u}{m_d}\,,
\label{eq:QCD-mass}
\end{align}
with $m_a^{\rm QCD}$, $f_\pi$, $m_\pi$, $m_u$ and $m_d$ denoting the QCD axion mass, the pion decay constant and the pion, up and down quark masses respectively. Since $\Lambda_{\rm QCD}\simeq 150 \,{\rm MeV}$ the mass is overall suppressed with respect to the generic expectation $m_a\sim\Lambda_{\rm QCD}^2/f_a$ in the absence of light charged fermions assumed in the previous section.\footnote{Note that alternative QCD axion scenarios have been proposed where the axion mass is suppressed~\cite{Hook:2018jle, DiLuzio:2021pxd, DiLuzio:2021gos} or enhanced, e.g.~\cite{Rubakov:1997vp,Berezhiani:2000gh,Hook:2014cda,Fukuda:2015ana,Agrawal:2017ksf,Gaillard:2018xgk,Csaki:2019vte}.}

Taking the above into account we can rewrite formula in \cref{Eq:simple ALP relic dansity ratio CMB} for the relic density in terms of the axion decay constant
\begin{align}
\frac{\rho_{a,0}}{\rho_{\rm DM}} \simeq \left(\frac{f_a}{1.84\cdot 10^{11}\, {\rm GeV}}\right)^{7/6} \frac{\mathcal{F}}{g_{\rm \rho,SM}(T_{\rm osc})^{1/6}}\,\,,
\label{eq:QCDDM}
\end{align}
where we used $T_c=\Lambda_{\rm QCD}$, $\beta=4$ and $\theta_i=1$. \cref{eq:QCDDM} implies that for $f_a < 1.8\cdot 10^{11}\, {\rm GeV}$ the relic abundance is too little to account for the observed dark-matter abundance while for $f_a > 1.8\cdot 10^{11}\, {\rm GeV}$ the relic abundance is too high. In the next few paragraphs we will investigate how thermal friction could in principle open up both regimes for the QCD axion. We will treat the two regimes separately in a manner that is analogous to the previous section.

\begin{figure}
    \centering
    \includegraphics[scale=0.45]{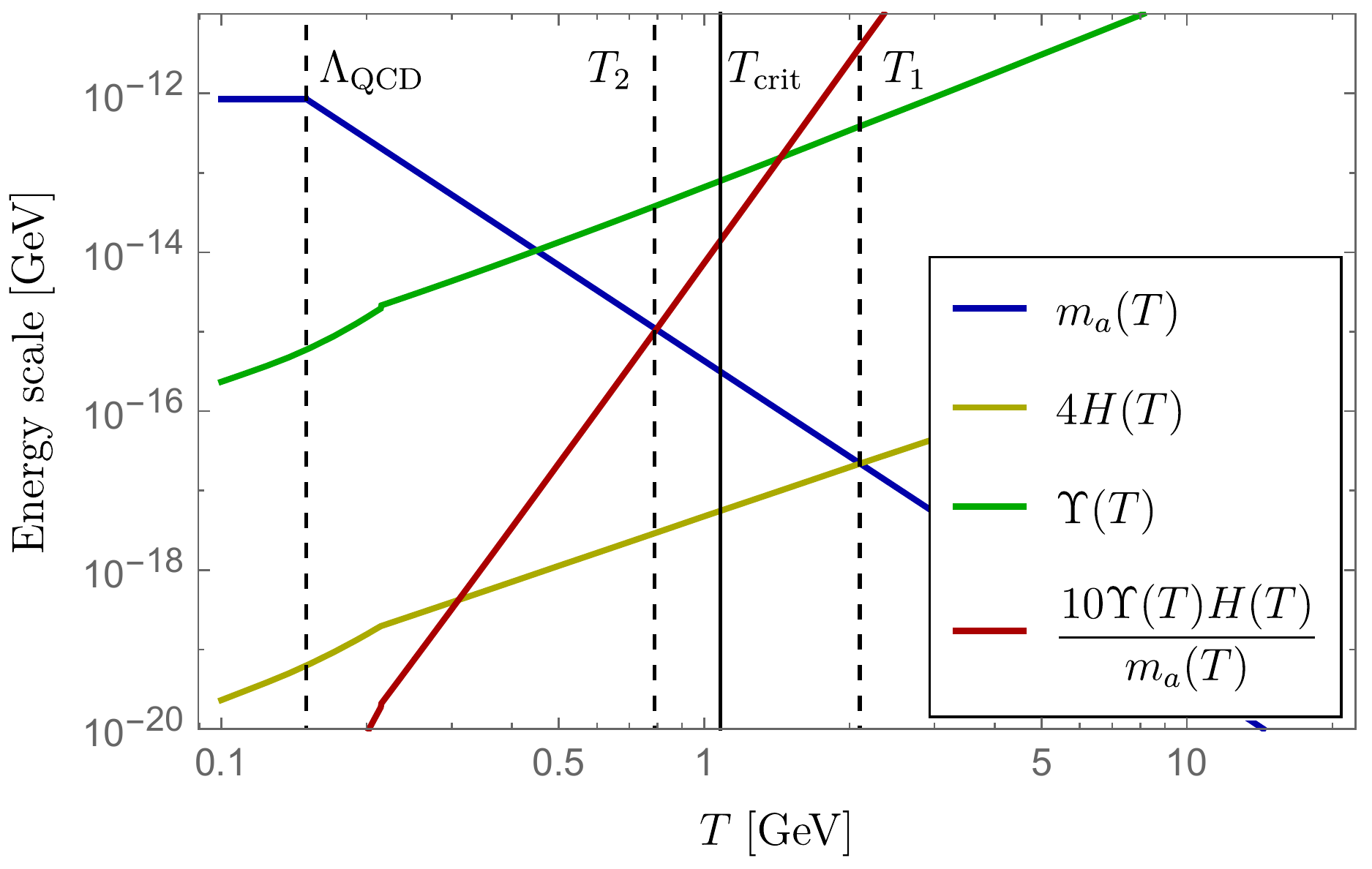}
    \caption{An example of the scenario that could open up the parameter space for QCD axion dark matter. The vertical lines represent the relevant scales defined in \cref{eq:temperature-conditionsALP} except for $\Lambda$ which is too low to be seen in the panel. Normally the onset of oscillations occurs at $T_1$, however the friction can delay the onset at most until $T_2$. If the dark sector spontaneously breaks at $T_{\rm crit}$ then the correct abundance is obtained. For this example we used $f_a=10^{10}\; {\rm GeV}$, $\lambda=10^5$ and $\Lambda=10^{-3}\; {\rm GeV}$.}
    \label{fig:QCD-example}
\end{figure}

\medskip
\noindent\textbf{Underabundant QCD Axion\,---\,}%
%
The standard expression for the relic abundance in \cref{Eq:simple ALP relic dansity ratio CMB} implicitly assumes that the oscillation temperature is the solution of the upper branch of \cref{eq:oscillation-temperature}. However, in the presence of thermal friction, the true temperature at which the onset of oscillations occurs is given by the lower branch of  \cref{eq:oscillation-temperature}. We may derive a general formula for the relic abundance given the parameters of the QCD axion and assuming that the moment of the onset of oscillations is a free parameter
\begin{align}
\frac{\rho_{a,0}}{\rho_{\rm DM}} \simeq \left(\frac{ m_a f_a\theta_i}{T_{\rm osc}^2}\right)^2\left(\frac{104\,{\rm GeV}}{T_{\rm osc}}\right)^3 \frac{\mathcal{F}}{g_{\rm \rho,SM}(T_{\rm osc})^{3/4}}\,.
\label{eq:QCDDM2}
\end{align}
Assuming a non fine-tuned initial misalignment angle $\theta_i\simeq 1$ it is apparent from \cref{eq:QCD-mass} that the product $m_a f_a$ is a constant associated to SM parameters which implies that there is one precise onset-of-oscillations temperature which guarantees that the QCD relic abundance will be the observed dark-matter one. This temperature is
\begin{align}
T_{\rm crit}\simeq 1.08\;{\rm GeV}\;\;\;\longleftrightarrow\;\;\; {\rm \rho_{a,0}\simeq\rho_{\rm DM}}\,.
\end{align}
In the presence of thermal friction, it is possible to delay the onset of oscillations until this critical temperature provided the friction is strong enough to prevent the axion from rolling until then. Considering all of the above we conclude that in order to open up the $f<1.8\cdot 10^{11}\, {\rm GeV}$ parameter space, we need the dark sector to be spontaneously broken at the temperature $T_{\rm crit}$. \cref{fig:QCD-example} displays a concrete example of the various relevant mass dimension quantities as functions of temperature.

Just as in the previous section, we require that the friction is strong enough so that the QCD axion remains frozen until at least $T_{\rm crit}$. This implies that
\begin{align}
\frac{1.04\cdot 10^{-6}\lambda^2}{\left[1+\ln\left(\frac{\Lambda_{\rm QCD}^2}{\Lambda^2}\right)/3.420\right]^5}>1\,.
\label{eq:ineq}
\end{align}

\begin{figure}
    \centering
    \includegraphics[width=0.45\textwidth]{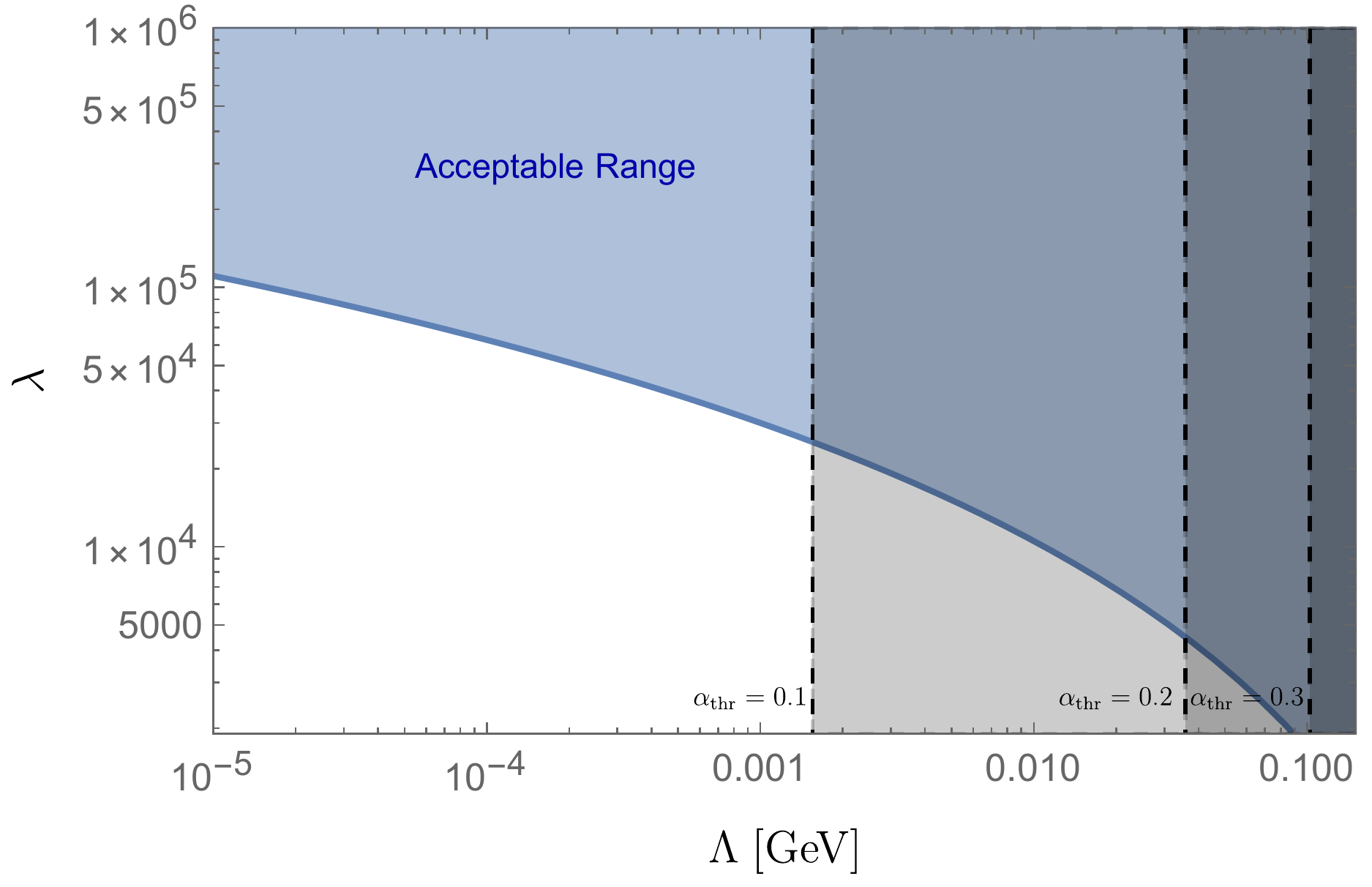}
    \caption{Region of the parameter space $\{\lambda,\Lambda\}$ that allows to reproduce the correct axion DM relic density.
    The blue shaded region ensures that the QCD axion does not roll until $T_{\rm crit}$ at which time the dark sector is spontaneously broken. Finally the gray shaded regions signal the parameter space for which our expressions for the thermal friction breakdown before $T_{\rm crit}$ is reached for different options of the threshold value of the running coupling.}
    \label{fig:QCDRange}
\end{figure}

In \cref{fig:QCDRange} the acceptable parameter space is displayed. It is important to note that this result is independent of the exact QCD axion mass or decay constant since it only depends on the SM parameters through the product $m_af_a$, see \cref{eq:QCD-mass}. It only assumes that $T_{\rm end}\simeq T_{\rm crit}$ and hence we consider the entire range $ 10^8\,{\rm GeV} < f_a < 1.8 \cdot 10^{11}\, {\rm GeV}$ to be viable.

\medskip
\noindent\textbf{Overabundant QCD Axion\,---\,}%
%
The main tools for computing the abundance and determining the required $\{\lambda,\Lambda\}$ to open up the parameter space were explained in detail in the previous section. Here we simply rewrite the relevant equations adapted for the QCD axion
\begin{align}
\frac{\rho_{a,0}}{\rho_{\rm DM}}&=2.4\cdot 10^{-3}\,{\rm e}^{-D}\left(\frac{\lambda^{14}}{\tau_{\rm 2}^{ 35}}\right)^{1/13}\label{eq:QCDoverRho}\,,\\
\tau_{\rm 2}&=\ln\left[\frac{1.97\,{\rm GeV}}{\Lambda}\left(\frac{\tau_{\rm 2}^{ 5}}{\lambda^2}\right)^{1/13}\right]\label{eq:ToscYQCD}\,,\\
\tau_{\rm 1}&=\ln\left(\frac{169\,{\rm GeV}^{5/6}\,m_a^{1/6}}{\Lambda}\right)\,,  \label{eq:ToscHQCD}
\end{align}
and the suppression parameter $D$ is given by the same expression as in the case of a generic ALP. Aside from replacing the equations above, the algorithm for determining acceptable values of $\{\lambda,\Lambda\}$ pairs remains unchanged. Some indicative choices are found in table \cref{table:2}.

\begin{table}[h!]
\centering
\begin{tabular}{|c| c| c| c| c| c| c|} 
 \hline
 No & $f_a$ (GeV) & $m_a$ (eV) & $\lambda$ & $\Lambda$ (GeV) & $\tau_{\rm 2}$ & $\tau_{\rm end}$ \\
 \hline
 1 & $10^{13}$ & $8.5\times 10^{-7}$ & $1.5\times10^{6}$ & $1\times 10^{-4}$ & 8.52 & 6.50 \\ 
 2 & $10^{15}$ & $8.5\times 10^{-9}$ & $4.3\times 10^{8}$ & $3\times 10^{-5}$ & 8.88 & 7.00 \\
 3 & $10^{17}$ & $8.5\times 10^{-11}$ & $1\times 10^{11}$ & $3\times10^{-6}$ & 10.40 & 6.50\\ 
 \hline
\end{tabular}
\caption{Sample values of $\{\lambda,\Lambda\}$ that induce sufficient friction to deplete the axion relic abundance to the observed one for a given choice of $\{m_a,f_a\}$ for the QCD axion.}
\label{table:2}
\end{table}

\section{Discussion and Conclusions}
\label{sec:conclusions}

The main aim of this work has been to extend the standard misalignment mechanism for the generation of axion dark matter in the presence of sphaleron-induced thermal gauge friction. The coupling of the axion to a dark non-abelian gauge sector in a secluded thermal bath can significantly impact the prediction for the axion abundance.
This "frictional misalignment" mechanism can result in an enhancement of the axion relic density by delaying the onset of oscillations or in a depletion of the abundance if the friction is active during the oscillations. Taking into account the one-loop running of the dark coupling constant, we have derived an analytical expression for the \emph{frictional adiabatic invariant} which can be used to compute the axion abundance in a variety of models since it is a constant of motion in the presence of thermal friction. 

We have then applied this general mechanism to some particular models of interest. In the most minimal case where a single non-abelian confining gauge group is responsible for both the ALP mass and the friction, we find that the prediction of axion dark matter departs from the standard result for masses $m_a\gtrsim 100\;{\rm eV}$. In order to obtain a reliable prediction of the abundance, further studies on the sphaleron rate close to the confinement scale are required.  
As a second example we studied an axion that couples instead to two separate gauge groups: one confines and generates the axion mass while the other is spontaneously broken and generates the friction. We find that, in order for the friction to have an impact, an enhancement of the coupling to the spontaneously broken group is required, which could arise from clockwork or alignment scenarios. In this case, the window for axion dark matter opens up and for different values of the enhancement parameter, both the traditionally under- and overabundant regions are populated within the frictional misalignment mechanism. Analogously, this also applies to the QCD axion.

Couplings between axions and gauge fields, in a cosmological setting, result in rich phenomenology that should be further studied and understood. Axion-gauge couplings may lead to tachyonic, non thermal enhancement of gauge modes which results in rich phenomenology in many aspects of cosmology such as inflation \cite{Anber:2009ua,Barnaby:2011vw,Domcke:2019lxq,Domcke:2020zez,Gorbar:2021rlt,Kitajima:2017peg}, dark matter \cite{Agrawal:2017eqm,Machado:2018nqk,Ratzinger:2020oct}, gravitational waves \cite{Machado:2019xuc,Ratzinger:2020koh,Madge:2021abk}, cosmological relaxation \cite{Hook:2016mqo,Domcke:2021yuz,Ji:2021mvg,Banerjee:2021oeu} or dark energy \cite{DallAgata:2019yrr}. On the other hand, the cosmological implications of thermal backgrounds of gauge fields have only been recently explored  in a limited setting and mainly in the context of inflation and not axion dark matter ({see also \cite{Ferreira:2017lnd,Ferreira:2017wlx,Ji:2021mvg,DeRocco:2021rzv} for the transition from the non-thermal to thermal regime}). Our work fills this gap in the literature and provides a natural extension to the standard misalignment mechanism that is based on a set of reasonable assumptions which however have a strong impact on the size of the viable parameter space for axion dark matter. 


\medskip\noindent\textit{Acknowledgments\,---\,}%
A.~P.\ and P.~Q.\ would like to thank the CERN Theory Division for the warm hospitality where this work was initiated.
This work was supported by IBS under the project code, IBS-R018-D1.
P.~Q.\ acknowledges support by the Deutsche Forschungsgemeinschaft under Germany's Excellence Strategy\,--\,EXC 2121 Quantum Universe\,--\,390833306 and 491245950.
This project has received funding and support from the European Union's Horizon 2020 research and innovation programme under the Marie Sk\l odowska-Curie grant agreements No.\ 860881-HIDDeN and 796961 ("AxiBAU"). The work of P.Q. is supported in part by the U.S. Department of Energy Grant
No. DE-SC0009919.

\medskip\noindent{\bf Note added:} Shortly after the submission of the first version of our paper, Ref.~\cite{Choi:2022nlt} appeared on the arXiv, which also studies the production of axion dark matter in the presence of thermal friction. The discussion of the traditionally overabundant scenario largely overlaps with our work, while the traditionally underabundant scenario and the minimal ALP model are not addressed. We note that our work was initiated at CERN in summer 2021, in the context of the \textit{Axion Young Talent} initiative organized by one of us (K.~S.). First preliminary results of our work were shared on January 27th, 2022 at a closed journal club presentation at IBS-CTPU.



\bibliographystyle{JHEP}
\bibliography{arxiv_1}

\onecolumngrid
\appendix

\newpage

\section[]{Sphaleron rate for a generic $SU(N_c)$ gauge theory}
\label{App:SphlaeronValerie}
The sphaleron rate per unit time-volume in a $SU(N_c)$ gauge theory with $N_H$ complex scalars and no fermions reads ~\cite{Bodeker:1999gx,Arnold:1999ux,Arnold:1999uy,Moore:2000mx,Moore:2000ara,Moore:2010jd},
\begin{align}
2\, \Gamma_{\text {sphal }} &\simeq 0.21\left(\frac{N_{c} g^{2} T^{2}}{m_{D}^{2}}\right)\left(\ln \frac{m_{D}}{\gamma}+3\right) \frac{N_{c}^{2}-1}{N_{c}^{2}}\left(N_{c} \alpha\right)^{5} T^{4}\,, \\
\gamma&=N_{c} \alpha T\left(\ln \frac{m_{D}}{\gamma}+3.041\right) \,,\\
m_{D}^{2}&=\frac{2 N_{c}+N_{H}}{6} g^{2} T^{2}\,,
\end{align}
where $g\equiv \sqrt{4 \pi \alpha}$. For a given choice of the fundamental parameters the sphaleron rate needs to be evaluated recursively and numerically. For our purposes it will be enough to approximate it as follows,
\begin{align} 
   \Gamma_{\rm sph}\simeq1.8\times\frac{N_c^2-1}{N_c^2}{\left(N_c \alpha\right)^5 T^4}\,.
\end{align}

\section{Analytical derivation of the frictional adiabatic invariant}
\label{sec:analytical_derivation_of_the_frictional_invariant}
Using the WKB approximation one can obtain the new frictional adiabatic invariant
\begin{align}
\frac{\rho_\theta a^3}{m_a(T)} {\rm e}^{\int^{t}_{t_0} \Upsilon\left[T'(\tilde{t})\right] d\tilde{t}} ={\rm cte}\,,
\end{align}
where the friction reads
\begin{equation}
    \Upsilon(T')=C_{\rm new}\frac{\left(N_c \alpha\right)^5 T'^3}{2  f_a^2}\,.
\end{equation}
Defining $C_{\rm new}\equiv1.8\times\frac{N_c^2-1}{N_c^2}$ and parametrizing  the running of $\alpha$ as,
\begin{align}
\alpha\left(T'\right)=
\frac{4\pi}{\bar{b}_0 N_c}\frac{1}{\log\left(\frac{T'^2}{\Lambda^2}\right)}=
 \frac{4\pi}{\bar{b}_0 N_c}\frac{1}{2\tau}\,, \qquad \text{ with } \qquad \tau\equiv \log\left(\frac{T'}{\Lambda}\right)\,,
\end{align}
one obtains, 
\begin{equation}
    \Upsilon(\tau)=
     C_{\rm new} \frac{ 2^4 \Lambda^3 {\pi^5}}{\bar{b}_0^5 \, f^2} \, \frac{e^{3\tau}}{\tau^5 }\,,
\end{equation}
and let's remember that,
\begin{align}
T=\xi^{-1} \left(\frac{g_{\rm s, SM}(T_0)\;g'_{\rm s}(T')}{g_{\rm s, SM}(T)\;g'_{\rm s}(T'_0)}\right)^{1/3} T'\,.
\end{align}
The goal is to obtain the exponent by performing the integral
\begin{align}
D =\int^{t}_{t_0} \Upsilon\left[T'(\tilde{t})\right] d\tilde{t}=
 \int^{\tau_{\rm end}}_{\tau_{\rm osc}} \Upsilon \left(T'\right) \:\frac{dt}{dT}\frac{dT}{dT'} \frac{dT'}{d\tau}d\tau\,\,,
\end{align}
where we have conveniently made some changes of variables whose jacobians read:
\begin{align}
\frac{d t}{d T}&=-M_p \sqrt{\frac{45}{8 \pi^{2}}} \frac{1}{T^{3} g_{s}(T) \sqrt{g_{\rho}(T)}}\left(T \frac{d g_{\rho}(T)}{d T}+4 g_{\rho}(T)\right) 
\simeq 
-\frac{4 M_p}{T^{3}} \sqrt{\frac{45}{8 \pi^{2}}} \frac{ \sqrt{g_{\rho}(T)}}{ g_{s}(T) }\,,\\
\frac{dT}{dT'}&\simeq 
\xi^{-1}\left(\frac{g_{\rm s, SM}(T_0)\;g'_{\rm s}(T')}{g_{\rm s, SM}(T)\;g'_{\rm s}(T'_0)}\right)^{1/3} \,,\\
\frac{dT'}{d\tau}&=\Lambda\, e^\tau\,,
\end{align}
where we neglect $d g_{\rho,s}(T)/{d T}$. Putting everything together we obtain,
\begin{align}
D =
 \int^{\tau_{\rm end}}_{\tau_{\rm osc}}
  C_{\rm new} \frac{ 2^4\Lambda^3 {\pi^5}}{\xi\,\bar{b}_0^5  f_a^2} \, \frac{e^{3\tau}}{\tau^5 } 
  \left[-{4 M_p} \sqrt{\frac{45}{8 \pi^{2}}} \frac{ \sqrt{g_{\rho, \text{ SM}}(T)}}{ g_{s, \text{ SM}}(T) }\right] 
  \left(\frac{g_{\rm s, SM}(T)\;g'_{\rm s}(T'_0)}{g_{\rm s, SM}(T_0)\;g'_{\rm s}(T')}\right)^{2/3}
  \frac{1}{\Lambda^2\, e^{2\tau}}
  d\tau\,.
\end{align}
Assuming a mild dependence on the variation of the degrees of freedom, we factor them out of the integral choosing their mean values defined as
\begin{equation}
    \bar{g}=\frac{1}{T_{\rm end}-T_{\rm osc}}\int^{T_{\rm end}}_{T_{\rm osc}}g(T)\,dT\,.
\end{equation}
For the degrees of freedom of the dark thermal bath we simply evaluate the function at $T_{\rm osc}$ for concreteness. The exact temperature does not matter since the function is constant until confinement/spontaneous breaking of the dark sector. Our result now reads
\begin{align}
D= -\frac{\Lambda M_p}{f_a^2} C_{\rm new} 
\frac{ 2^6 {\pi^5}}{\xi\,\bar{b}_0^5 } \,  \sqrt{\frac{45}{8 \pi^{2}}} 
  \left(\frac{g'_{\rm s}(T'_0)}{g_{\rm s, SM}(T_0)\;g'_{\rm s}(T_{\rm osc})}\right)^{2/3}
  \frac{ \sqrt{\bar{g}_{\rho, \text{ SM}}}}{\bar{g}_{s, \text{ SM}}^{1/3} }
  \int^{\tau_{\rm end}}_{\tau_{\rm osc}}\frac{e^{\tau}}{\tau^5 } d\tau\,.
\end{align}
Finally the only integral we need to compute is
\begin{align}
\int^{\tau_{\rm end}}_{\tau_{\rm osc}} \frac{e^{\tau }}{\tau ^5} \, d\tau=
\frac{1}{24} \left[\text{Ei}(\tau )-\frac{e^{\tau } \left(\tau ^3+\tau ^2+2 \tau +6\right)}{\tau ^4}\right]^{\tau_{\rm end}}_{\tau_{\rm osc}}\,,
\end{align}
where ${\rm Ei}(z)=-\int^\infty_{-z} {\rm e}^{-t}/t \;dt$ is the Exponential Integral function. 

Our final result reads
\begin{align}
D= -\frac{8\Lambda M_p}{3 f_a^2} C_{\rm new} 
\frac{ {\pi^5}}{\xi\,\bar{b}_0^5 } \,  \sqrt{\frac{45}{8 \pi^{2}}} 
  \left(\frac{g'_{\rm s}(T'_0)}{g_{\rm s, SM}(T_0)\;g'_{\rm s}(T_{\rm osc})}\right)^{2/3}
  \frac{ \sqrt{\bar{g}_{\rho, \text{ SM}}}}{\bar{g}_{s, \text{ SM}}^{1/3} } \left[\text{Ei}(\tau )-\frac{e^{\tau } \left(\tau ^3+\tau ^2+2 \tau +6\right)}{\tau ^4}\right]^{\tau_{\rm end}}_{\tau_{\rm osc}}\,.
\end{align}

\paragraph*{Mild running approximation}

For large temperatures $T'\gg \Lambda$ the running of the coupling constant is mild and the expression of the frictional misalignment can be further simplified,
\begin{align}
\int^{\tau_{\rm end}}_{\tau_{\rm osc}} \frac{e^{\tau }}{\tau ^5} \, d\tau=
\frac{1}{24} \left[\text{Ei}(\tau )-\frac{e^{\tau } \left(\tau ^3+\tau ^2+2 \tau +6\right)}{\tau ^4}\right]^{\tau_{\rm end}}_{\tau_{\rm osc}}\xrightarrow[]{\tau\rightarrow \infty} \left[\frac{e^{\tau }}{\tau^5}\right]^{\tau_{\rm end}}_{\tau_{\rm osc}} =\left[\frac{T'/\Lambda}{\left[\ln(T'/\Lambda)\right]^5}\right]^{T'_{\rm end}}_{T'_{\rm osc}}\,,
\end{align}
leading to
\begin{align}
D= -\frac{\Lambda M_p}{f_a^2} C_{\rm new} 
\frac{  2^6 {\pi^5}}{\bar{b}_0^5 } \,  \sqrt{\frac{45}{8 \pi^{2}}} 
  \left(\frac{g'_{\rm s}(T'_0)}{g_{\rm s, SM}(T_0)\;g'_{\rm s}(T_{\rm osc})}\right)^{2/3} \frac{ \sqrt{\bar{g}_{\rho, \text{ SM}}}}{\bar{g}_{s, \text{ SM}}^{1/3} } \left[\frac{T'/\Lambda}{\left[\ln(T'/\Lambda)\right]^5}\right]^{T'_{\rm end}}_{T'_{\rm osc}}\,.
\label{Eq:Dsimpl}
\end{align}
\newpage
\end{document}